\documentclass[fleqn,usenatbib,useAMS]{mnras}

\usepackage{graphicx}	 % Including figure files
\usepackage{amsmath}	  % Advanced maths commands
\usepackage{amssymb}	  % Extra maths symbols
\usepackage{multicol}  % Multi-column entries in tables
\usepackage{bm}		      % Bold maths symbols, including upright Greek
\usepackage{pdflscape}	% Landscape pages
\usepackage{hyperref}
\usepackage{xcolor}   

\newcommand{\HII}{\mbox{H\,{\sc ii}}}
\newcommand{\HeI}[1]{\mbox{He\,{\sc i}~$\lambda${#1}}}
\newcommand{\HeII}[1]{\mbox{He\,{\sc ii}~$\lambda${#1}}}

\newcommand{\OIt}[1]{\mbox{O\,{\sc i}~$\lambda\lambda\lambda${#1}}}

\newcommand{\NaId}[1]{\mbox{Na\,{\sc i}~$\lambda\lambda${#1}}}
\newcommand{\KI}[1]{\mbox{K\,{\sc i}~$\lambda${#1}}}
\newcommand{\KId}[1]{\mbox{K\,{\sc i}~$\lambda\lambda${#1}}}

\newcommand{\GBP}{\mbox{$G_{\rm BP}$}}
\newcommand{\GRP}{\mbox{$G_{\rm RP}$}}

\newcommand{\mcii}[1]{\multicolumn{2}{c}{#1}}
\newcommand{\mlii}[1]{\multicolumn{2}{l}{#1}}

\newcommand{\EBV}{\mbox{$E(4405-5495)$}}
\newcommand{\RV}{\mbox{$R_{5495}$}}
\newcommand{\AV}{\mbox{$A_{5495}$}}
\newcommand{\EWb}{\mbox{EW$_{7700}$}}
\newcommand{\EWKIr}{\mbox{EW$_{{\rm K\,I\,} \lambda 7701.093}$}}
\newcommand{\chired}{\mbox{$\chi^2_{\rm red}$}}

\usepackage[T1]{fontenc}
\usepackage{ae,aecompl}

\usepackage{newtxtext,newtxmath}
\title[Galactic extinction laws: II. A broad ISM absorption band]{Galactic extinction laws: II. Hidden in plain sight, a new \linebreak
                                                                 interstellar absorption band at 7700~\AA\ broader than any known DIB} % REFEREE

\author[J. Ma{\'\i}z Apell\'aniz et al.]{J. Ma{\'\i}z Apell\'aniz,$^{1}$\thanks{E-mail: \href{mailto:jmaiz@cab.inta-csic.es}{jmaiz@cab.inta-csic.es}}
R. H. Barb\'a,$^{2}$
J. A. Caballero,$^{1}$
R. C. Bohlin$^{3},$
and 
C. Fari\~na$^{4,5}$
\\
$^{1}$Centro de Astrobiolog\'{\i}a. CSIC-INTA. Campus ESAC. Camino bajo del castillo s/n. E-28\,692 Villanueva de la Ca\~nada. Madrid. Spain.\\
$^{2}$Departamento de Astronom\'{\i}a. Universidad de La Serena. Av. Cisternas 1200 Norte. La Serena. Chile.\\
$^{3}$Space Telescope Science Institute. 3700 San Martin Drive. Baltimore, MD 21\,218, U.S.A.\\
$^{4}$Instituto de Astrof{\'\i}sica de Canarias. E-38\,200 La Laguna, Tenerife, Spain.\\
$^{5}$Isaac Newton Group of Telescopes. Apartado de correos 321. E-38\,700 Santa Cruz de La Palma, La Palma, Spain.\\
}

\date{Accepted 2020 August 3. Received 2020 August 3; in original form 2020 June 4.}

\pubyear{2020}

\begin{document}
\label{firstpage}
\pagerange{\pageref{firstpage}--\pageref{lastpage}}
\maketitle

\begin{abstract}
We have detected a 
broad interstellar absorption % REFEREE
band centred close to 7700~\AA\ and with a FWHM of 176.6$\pm$3.9~\AA. This is the first such absorption band detected in the optical 
range and is significantly wider than the numerous diffuse interstellar bands (DIBs). It remained undiscovered until now because it is partially hidden behind the A telluric band produced
by O$_2$. The band was discovered using STIS@HST spectra and later detected in a large sample of stars of diverse type (OB stars, BA supergiants, red giants) using further STIS and
ground-based spectroscopy. The EW of the band is measured and compared with our extinction and \KId{7667.021,7701.093} measurements for the same sample. The carrier is ubiquitous in the
diffuse and translucent Galactic ISM but is depleted in the environment around OB stars. In particular, it appears to be absent or nearly so in sightlines rich in molecular carbon. This 
behaviour is similar to that of the $\sigma$-type DIBs, which originate in the low/intermediate-density UV-exposed ISM but are depleted in the high-density UV-shielded molecular clouds. We 
also present an update on our previous work on the relationship between \EBV\ and \RV\ and incorporate our results into a general model of the ISM.
%and discuss some aspects of the detailed behaviour of the extinction law with wavelength that will be analysed in future papers of this series.
\end{abstract}


%%%%%%%%%%%%%%%%%%%%%%%%%%%%%%%%%%%%%%%%%%%%%%%%%%

%%%%%%%%%%%%%%%%% BODY OF PAPER %%%%%%%%%%%%%%%%%%

\section{Introduction}

$\,\!$\indent The ISM imprints a large number of absorption lines in the optical range of stellar spectra. A few lines are well identified and their origin ascribed 
to atomic (e.g. \NaId{5891.583,5897.558})\footnote{All wavelengths in this paper are given in \AA\ in vacuum.} or molecular (e.g. CH~$\lambda$4301.523) species but many of 
them are grouped under the term ``diffuse interstellar bands'' (DIBs) and their origin is debated, with just a few of them having had their carriers identified
\citep{Campetal15,Cordetal19}. DIBs were discovered almost one century ago by \citet{Hege22} and the ``diffuse'' in the name refers to their larger intrinsic widths 
compared to atomic and molecular lines. They are typically divided into narrow (FWHM around 1~\AA) and broad (FWHM from several \AA\ up to 40~\AA, e.g. 
\citealt{Maizetal15c}).

In the UV the situation is different, with many atomic ISM lines 
but no confirmed DIBs \citep{SeabSnow85,Clayetal03b}, % REFEREE
and the most significant absorption structure is the broad 2175~\AA\ absorption band, 
which is much wider and stronger than 
any optical DIB \citep{BlesSava70,Sava75}. % REFEREE
In the near infrared no broad absorption bands have been reported (but a few 
weak DIBs have) and in the mid infrared several broad absorption bands are seen and have been ascribed to silicates, H$_2$O, CO, CO$_2$, and aliphates (\citealt{Fritetal11} and 
references therein).

In the optical range no clearly defined absorption bands have been reported until now but several very broad band structures have been detected with
sizes of $\sim$1000~\AA\ and larger \citep{Whit58,Whit66,York71,Massetal20}, % REFEREE
sometimes qualifying them as ``knees'' in the extinction law.
As most previous families of extinction laws (e.g. \citealt{Cardetal89,Fitz99,Maizetal14a}) 
are based on broad-band photometric data, those structures have been included to some extent in their analysis. However, \citet{Cardetal89}
used a polynomial interpolation in wavelength using as reference the Johnson filters and this led to artificial structures and incorrect results when using
Str\"omgren photometry. The issue was addressed by \citet{Maizetal14a} by switching to spline interpolation. 

This is the second paper of a series on Galactic extinction laws (but as \citealt{MaizBarb18} is extensively cited in the text, it will be referred to
as paper~0) 
that plans to cover the UV-optical-IR ranges. In paper I \citep{Maizetal20a} % REFEREE
we analysed the behaviour of extinction laws in the NIR using 2MASS data. Here we report the discovery of the first 
optical broad 
(FWHM $>$ 100~\AA, broader than broad DIBs, see above) % REFEREE
absorption band, centred at 7700~\AA. We start describing the discovery of the new absorption
band, then discuss the methods used to measure it, continue with our results and analysis, and finish with a summary and future plans. As the 7400--8000~\AA\ wavelength range has been 
seldom explored spectroscopically relative to other parts of the optical spectrum, we include two appendices with the ISM and stellar lines that we have found in the data.

\section{How did we find the new absorption band?}

$\,\!$\indent Our first encounter with the new absorption band took place when preparing the spectral library for \citet{MaizWeil18}. In that paper we produced a new photometric
calibration for {\it Gaia}~DR2 that was needed because of the presence of small colour terms when one compared the existing calibrations with good-quality spectrophotometry. Existing
libraries that had been used for previous calibrations such as NGSL \citep{Maiz17a} had different issues so we built a new stellar library combining {\it Hubble Space Telescope} 
({\it HST}) Space Telescope Imaging Spectrograph (STIS) G430L+G750L wide-slit low-resolution spectroscopy from different sources, mostly CALSPEC \citep{Bohletal19}, 
HOTSTAR \citep{KhanWort18}, and the Massa library \citep{Massetal20}. When analysing the new spectral library, we noticed a broad absorption band close to 7700~\AA\ in the 
spectra of extinguished OB stars. The intensity of the band was correlated with the amount of extinction and was much broader than any of the DIBs analysed in that
region or indeed any DIBs (see e.g. \citealt{JennDese94,Hobbetal08,Hobbetal09,Maizetal15c}). 

The \citet{MaizWeil18} stellar library includes only stars with weak extinction ($\EBV \lesssim 1$, see \citealt{Maiz04c,Maiz13b} to understand why monochromatic quantities 
should be used to define the amount or type of extinction), so the absorption band was relatively weak in those stars. We have recently started {\it HST} GO program 15\,816 to extend 
the CALSPEC library to very red stars and in that way improve the {\it Gaia} photometric calibration for objects with large \GBP$-$\GRP, which are mostly high-extinction red giants 
(see paper I). In the first visit of that program we observed two such extinguished red giants (2MASS~J16140974$-$5147147 and 2MASS~J16141189$-$5146516)
and we easily detected the absorption band there with the same central wavelength and 
similar FWHM, but clearly deeper. This confirmed the reality of the absorption band and its association with interstellar extinction.

\begin{table}
\caption{Spectrographs used, sample size observed with each, and associated \EWb\ random uncertainties.}
\label{spectrographs}
\centerline{
\begin{tabular}{lcccccc}
telesc. /  & aper. & spectrograph & spectral & type      & \# & unc.  \\
observ.    & (m)   &              & resol.   &           &    & (\AA) \\
\hline
{\it HST}  & 2.4   & STIS         &  800     & long slit & 26 & 0.4   \\
LT         & 2.0   & FRODOspec    &  2200    & IFU       & 30 & 0.6   \\
%           &       &              &  5300    & IFU       &    &       \\
CAHA       & 3.5   & CARMENES     &  95\,000 & \'echelle & 26 & 0.6   \\
La Silla   & 2.2   & FEROS        &  46\,000 & \'echelle & 55 & 1.8   \\
NOT        & 2.5   & FIES         &  25\,000 & \'echelle & 49 & 2.0   \\
           &       &              &  67\,000 & \'echelle &  7 & 2.0   \\
Mercator   & 1.2   & HERMES       &  85\,000 & \'echelle & 44 & 0.9   \\
%VLT        & 8.2   & UVES         &  60\,000 & \'echelle &    &       \\
%           &       &              & 110\,000 & \'echelle &    &       \\
\hline
\end{tabular}
$\;\;\;\;$}
\end{table}

Why was not this absorption band found earlier? The main reason is that its short-wavelength side is located in the deep O$_2$ telluric A band (7595--7715~\AA), with the
strongest absorption in the short-wavelength part of that range (a much weaker H$_2$O telluric band is present around 7900~\AA). That is why the absorption band was 
identified in the few existing {\it HST} spectra rather than in the much more common ground-based data. Note that \citet{Massetal20} included this region in their study and that some 
of their stars overlap with our sample, clearly showing the absorption band but the authors missed it. It also appears in Fig.~7 of \citet{Calletal20} but was not identified as an
ISM absorption band there. Once we knew where to look it was not difficult to find the absorption band in 
ground-based data. For that purpose we used two types of spectra of OB stars and A supergiants. The first type is \'echelle data from four spectrographs (CARMENES@CAHA-3.5m, 
FEROS@La~Silla~2.2~m, FIES@NOT, and HERMES@Mercator, see Table~\ref{spectrographs}) that we have collected over the years for the LiLiMaRlin project 
({\bf Li}brary of {\bf Li}braries of {\bf Ma}ssive-Star High-{\bf R}eso{\bf l}ut{\bf i}on Spectra, \citealt{Maizetal19a})
with the purposes of studying spectroscopic binaries (MONOS, {\bf M}ultiplicity {\bf O}f {\bf N}orthern {\bf O}-type {\bf S}pectroscopic Systems, \citealt{Maizetal19b}) 
and the intervening ISM (CollDIBs, {\bf Coll}ection of {\bf DIBs}, \citealt{Maiz15a}). We also analyzed
data from two other spectrographs: CAF\'E@CAHA-2.2~m and UVES@VLT but ended up not using them. In the first case we discarded it because of the existence of several gaps between 
orders in the wavelength region of interest and in the second one because of the poor order stitching in most spectra.
The second type of spectra was acquired with the dual-beam FRODOspec spectrograph at the Liverpool Telescope (LT) for the Galactic O-Star
Spectroscopic Survey (GOSSS, \citealt{Maizetal11}) at a spectral resolution of 2200 at 7700~\AA.  The band is seen in both the \'echelle and the single-order spectra, especially 
when one excludes the wavelength region where the telluric absorption is strongest. The comparison between the two types of spectra produced an insight into a possible 
additional reason why the absorption band had not been discovered earlier. \'Echelle spectrographs are 
commonly used nowadays due to their high dispersion, which in our case is useful because it allows us to more easily subtract the contribution from the telluric, stellar 
(see Appendix B), and ISM (e.g \KId{7667.021,7701.093}) lines that are present in the data. However, the high-resolution comes with a problem: broad-wavelength structures 
such as the one we are interested in are dispersed in two or more orders and the blaze-function correction can introduce effects at the 1--2\% level that can bias results or 
even mask the structure altogether. In this respect, single-order spectrographs such as FRODOspec have the advantage of permitting a more accurate spectral rectification at the expense 
of a lower spectral resolution. As explained in the next section, our combination of both techniques has allowed us to study the new absorption band in a large sample of stars.

% Pipelines     executed to find band:    CARMENES, FEROS, Mercator, FIES new, LT
% Pipelines NOT     used to find band:    CAFÉ-BEANS (order gaps but executed), UVES (order stitching)
% Pipelines NOT executed to find band:    Stella (weak S/N)
% Spectrographs that do not reach region: FIES old, HET, OHP, HARPS
% Possible additional spectrographs:      X-shooter

\section{Methods}

$\,\!$\indent In the previous section we described how we discovered the new absorption band. Here we describe how we selected our sample and how we have measured the band. 
The measurement process takes a different (and more complicated) route because of the different properties of the spectroscopy we use (space- or ground-based; low, 
intermediate or high spectral resolution; single-order or \'echelle), the different types of stars present in the sample (mostly hot stars but also two red giants), and the need 
to take into account the different narrow lines present (telluric, stellar, and ISM). 

\subsection{Getting rid of those pesky absorption and emission lines}

$\,\!$\indent Our spectra 
contain % REFEREE
not only the absorption band we are interested in but also other ISM absorption, stellar absorption and emission, and (for ground-based data)
telluric absorption lines. With respect to the first two types we provide two appendices where we list our results for them. Those results are not the main topic of this paper 
but undoubtedly are of interest for other pursuits, especially considering that this wavelength region has been studied in less detail than others in the optical. 

We start with the telluric lines. First, we modified our pipelines for both LiLiMaRlin (high resolution) and LT (intermediate resolution) to rectify the spectra without using any
points in the region around the absorption band. Both pipelines automatically fit O$_2$ and H$_2$O telluric absorption in
this region using the \citet{Gardetal13} models and adjusting the spectral resolution, intensity, and velocity. As the O$_2$ lines are very deep, the central part of their cores
can be difficult to be properly fit in the high-resolution data. However, as for many of our stars we have more than one epoch and as the telluric lines change in position with 
respect to the astrophysical ones throughout the year, in those cases we combine different observations of the same target selecting in each case the wavelength regions that are 
less affected by the telluric absorption. Nevertheless, the wavelength region with the highest telluric O$_2$ absorption (7595--7650~\AA) cannot be properly corrected in most 
cases so we exclude it from our analysis.

With respect to the ISM lines there are several DIBs (see Appendix A) that can be ignored as they either fall outside the critical region of 7650--7750~\AA\ where the absorption band
is deeper or they are too weak to contribute significantly to our measurements. However, the \KId{7667.021,7701.093} doublet falls in the critical region and needs to be 
accounted for. To correct it, we have measured the two lines in the high resolution data and used the result to subtract Gaussian profiles of the proper equivalent width, velocity, 
and spectral resolution in the low- and intermediate-resolution spectra (note that the lines are unresolved in those data). In the high resolution data we simply interpolate between 
the adjacent wavelengths to eliminate the \KId{7667.021,7701.093} doublet.

Finally, different stellar lines are seen as a function of spectral type (see Appendix B). The most ubiquitous strong stellar line in our data is \HeII{7594.844},
which appears in absorption for O stars of all subtypes and becomes stronger for the earlier subtypes. However, we can safely ignore it for most purposes and simply exclude it
in our fitting of the line in the {\it Gaia} DR2 STIS sample, as it falls away from the critical 7650--7750~\AA\ region. For the ground-based data most of the line is in any case
overwhelmed by the head of the O$_2$ telluric A band. A similar case is the \OIt{7774.083,7776.305,7777.527} triplet seen in absorption in the late-B and early-O supergiants in 
our sample, as we can also exclude that region from our fits for those stars. On the other hand, there are three lines seen in emission in early-to-mid O stars that fall 
in the critical region so for those targets we followed a process analogous to the one for the \KId{7667.021,7701.093} doublet to correct for their influence (but here fitting 
emission lines as opposed to absorption ones, see Appendix B for details). 

\subsection{Determination of the properties of the absorption band from the Gaia STIS samples}

\begin{figure}
\centerline{\includegraphics[width=\linewidth]{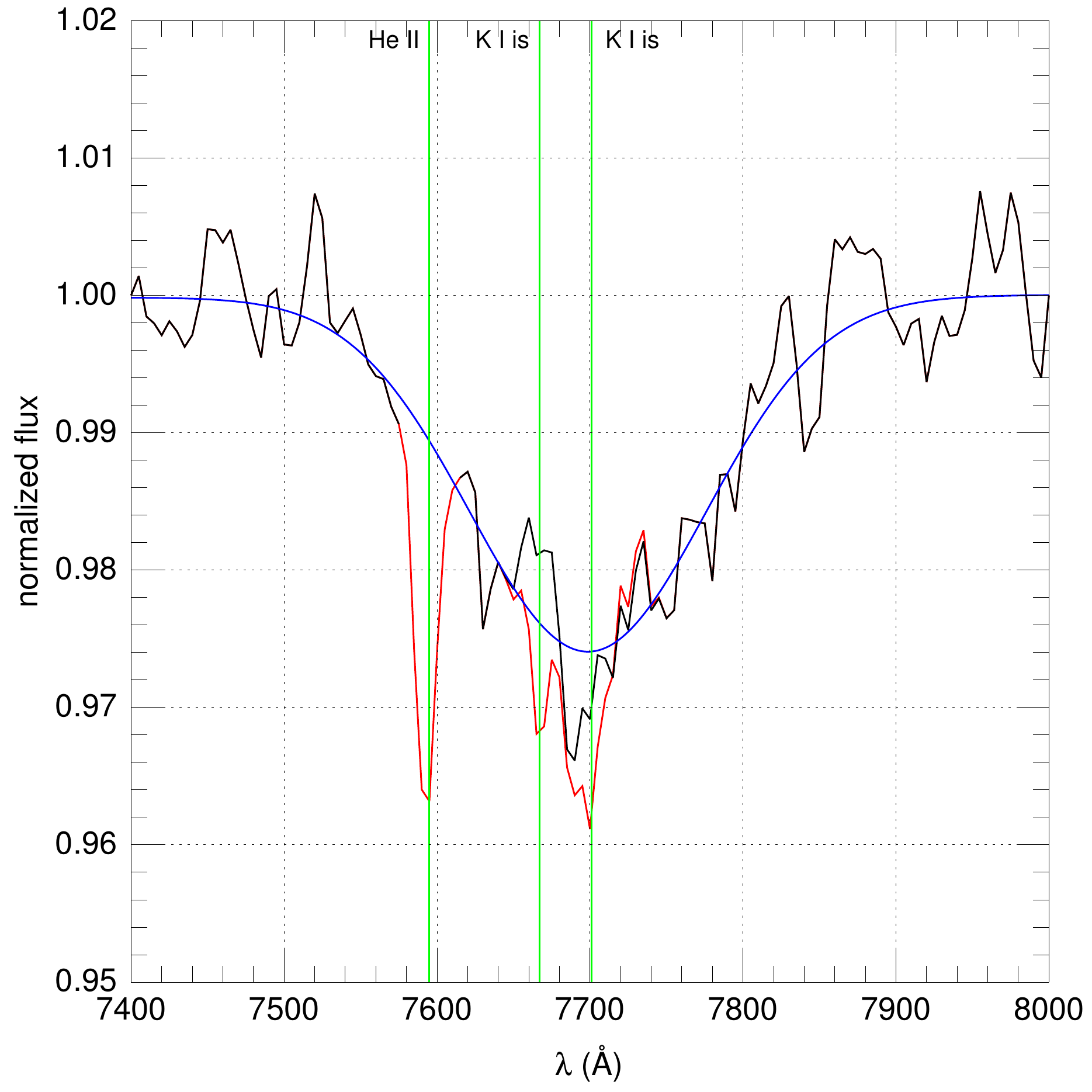}}
\caption{Combined weighted profile of the 7700~\AA\ band generated from 24 O4--B1-type G750L STIS spectra. The black line is the profile itself, the red line is the profile without
         subtracting the absorption/emission lines and without blocking the \HeII{7594.844} line, the blue line is the fitted Gaussian, and the green lines mark the position of the 
         three main absorption lines.}
\label{profile}
\end{figure}

$\,\!$\indent We combined the STIS samples we used for our analysis of {\it Gaia}~DR1 \citep{Maiz17a} and DR2 \citep{MaizWeil18} photometry, selecting the stars with O4--B1 spectral 
types from GOSSS, with good-quality NIR photometry, not 
affected by visual binarity, and for which we have good-quality high-resolution spectra. There are 24 such 
stars. For them we make an initial measurement of the absorption band by fitting an unrestricted Gaussian profile to each spectrum after (a) rectifying it, (b) correcting it for
emission/absorption lines as described in the previous subsection, and (c) shifting it to an ISM velocity frame as defined by the \KI{7701.093} velocity (the latter effect being 
very small due to the low spectral resolution of the STIS data). Then, a combined profile is obtained by merging the 24 spectra weighed by their equivalent widths
(Fig.~\ref{profile}). Finally, a Gaussian profile is fitted to the combined profile excluding the \HeII{7594.844} absorption line from the fit.

\begin{figure}
\centerline{\includegraphics[width=\linewidth]{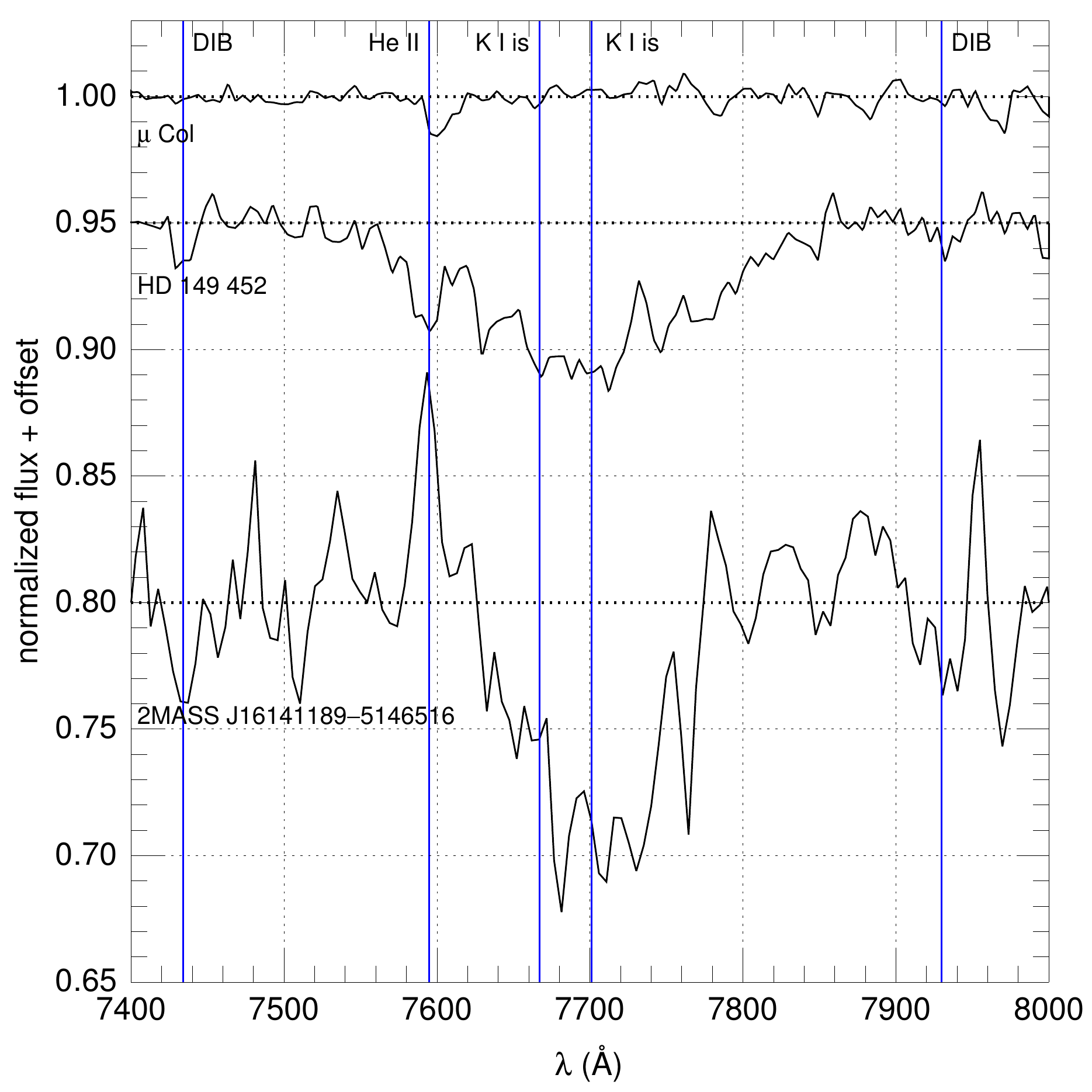}}
 \caption{Three sample G750L STIS spectra sorted by \EWb\ from non-detection to one of the red giants. The main stellar and ISM absorption lines have been left in place marked with
          blue vertical lines.}
\label{spectra_STIS}
\end{figure}

The Gaussian fit yields a central wavelength of 7699.2$\pm$1.3~\AA\ (vacuum) and a FWHM of 176.6$\pm$3.9~\AA, which we take as our preferred values for the absorption band. The fit
is of good quality within the S/N, which is not very high (Fig.~\ref{spectra_STIS})
because of the relatively short exposure times of the {\it HST} spectroscopy. We also fit a profile without subtracting the
emission/absorption lines (still excluding the \HeII{7594.844} absorption line from the fit) and in that case we obtain a central wavelength of 7695.1$\pm$1.1~\AA\ and a FWHM 
of 164.1$\pm$3.1~\AA, indicating that the effect of such lines (especially \KId{7667.021,7701.093}, as the emission lines are relatively weak in the STIS O4--B1 sample) introduces a 
significant bias if left uncorrected. The effect of not excluding the \HeII{7594.844} absorption line would be even greater, but in that case the profile would clearly deviate from 
a Gaussian. Therefore, from now on we refer to the absorption band by its approximate central wavelength of 7700~\AA.

Once we determined the FWHM of the profile, we fit a Gaussian profile of fixed width (see above) to each of the stars in the STIS O4--B1 sample to measure their equivalent widths (\EWb,
see below for the results). 
The main source of uncertainty for the EW in a relatively weak and broad absorption feature such as this one is the rectification of the spectrum. To determine it, we performed 
Monte Carlo simulations of the continuum around the line with the appropriate S/N and determined the scatter in the measured \EWb\ after rectifying each simulation, arriving at values 
between 0.3~\AA\ and 0.5~\AA. Therefore, we adopt an uncertainty of 0.4~\AA\ for the measurements of the STIS O4--B1 sample. See Fig.~\ref{spectra_STIS} for three sample spectra.

\subsection{Extending the sample to the high extinction regime}

$\,\!$\indent The O4--B1 STIS sample is relatively small and, more importantly, of low extinction (all objects have $\EBV < 1$~mag). As we are studying an ISM absorption feature, we need 
to extend our extinction range to higher values to properly study its behaviour. As already mentioned, we have done this with the two high-extinction red giants observed with STIS and by
extending our sample to 120 OBA stars with ground-based data and with a wide range of values of \EBV. The two red giants from {\it HST} GO program 15\,816 had to be treated differently from 
the OB stars observed with STIS (note they were not used to determine the profile in the previous subsection). We first estimated their temperature by comparing their rectified
spectra with models from the previously mentioned \citet{Maiz13a} grid for red giants obtained from the MARCS models of \citet{Gustetal08} and obtained a value of 3800~K.
As late-type stars have a multitude of absorption lines that cannot be seen at the low STIS resolution, we divided each spectra by that of the MARCS SED with the proper distance and 
extinction using the \citet{Maizetal14a} extinction laws (see below). For those two stars we also have to subtract the 
effect of the \KId{7667.021,7701.093} but we do not have high-resolution spectroscopy to do so. Therefore, we used the results from Appendix A to determine average EWs for 
\KId{7667.021,7701.093} for their extinctions and subtract the appropriate profiles in the STIS spectra. After doing that, we fit a Gaussian profile of fixed width as we previously 
did for the rest of the STIS sample and we obtain values of 14.0$\pm$1.3~\AA\ for 2MASS~J16140974$-$5147147 and 13.5$\pm$1.3~\AA\ for 2MASS~J16141189$-$5146516.

For the STIS sample we are limited by the relatively scarce sample available in the {\it HST} archive. For the ground-based data, on the other hand, we have a sample of over 1000 targets 
observed with FRODOspec and of 2000 targets in LiLiMaRlin to choose from. To build the sample we used several criteria:

\begin{itemize}
 \item Good S/N (Fig.~\ref{spectra_HR}) and, if possible, observed with two or three telescopes for cross-calibration purposes.
 \item Inclusion of all stars in the STIS O4--B1 sample.
 \item Targets with multiple epochs are favoured to facilitate the elimination of telluric lines.
 \item Objects with different values of extinction are included but preference is given to those with high \EBV\ or high \RV\ \citep{Maiz13a}, which are absent in the STIS O4--B1 sample.
 \item Most selected stars are of O spectral type but we have included some B stars that are of special interest as they have been previously used for ISM studies (e.g. Cyg~OB2-12 
       and HT~Sge) and one A supergiant (CE Cam).
\end{itemize}

As previously mentioned, in all ground-based spectra we first correct the telluric lines. Then, in the FRODOspec (intermediate resolution) spectra the stellar and ISM lines are corrected 
using Gaussian profiles and in the LiLiMaRlin (high resolution) the stellar lines are corrected in the same way and the \KId{7667.021,7701.093} doublet is interpolated from neighbour
wavelengths. Finally, we fit a Gaussian profile of fixed width blocking regions with strong telluric lines and defects.

\subsection{Cross-calibration of the data}                        

$\,\!$\indent Before using the values of \EWb\ band from the three different types of spectra in this paper (STIS long-slit low-resolution, FRODOspec IFU intermediate-resolution, and 
\'echelle high-resolution from different spectrographs) it is necessary to compare them to (a) identify and correct systematic biases between them and (b) properly characterize random 
uncertainties. In other words, to ensure that they are accurate and that we are using the right precision. 

We have already determined that the random uncertainty of the \EWb\ for the STIS O4--B1 sample is 0.4~\AA. We also judge that there is no systematic bias in those measurements for two 
reasons: the high quality of the spectrophotometric calibration of STIS in absolute terms \citep{Bohletal19} and compared to {\it Gaia}~DR2 \citep{MaizWeil18} and the fact that for the 
stars with negligible extinction we measure values of zero for \EWb. 

\begin{figure}
\centerline{\includegraphics[width=\linewidth]{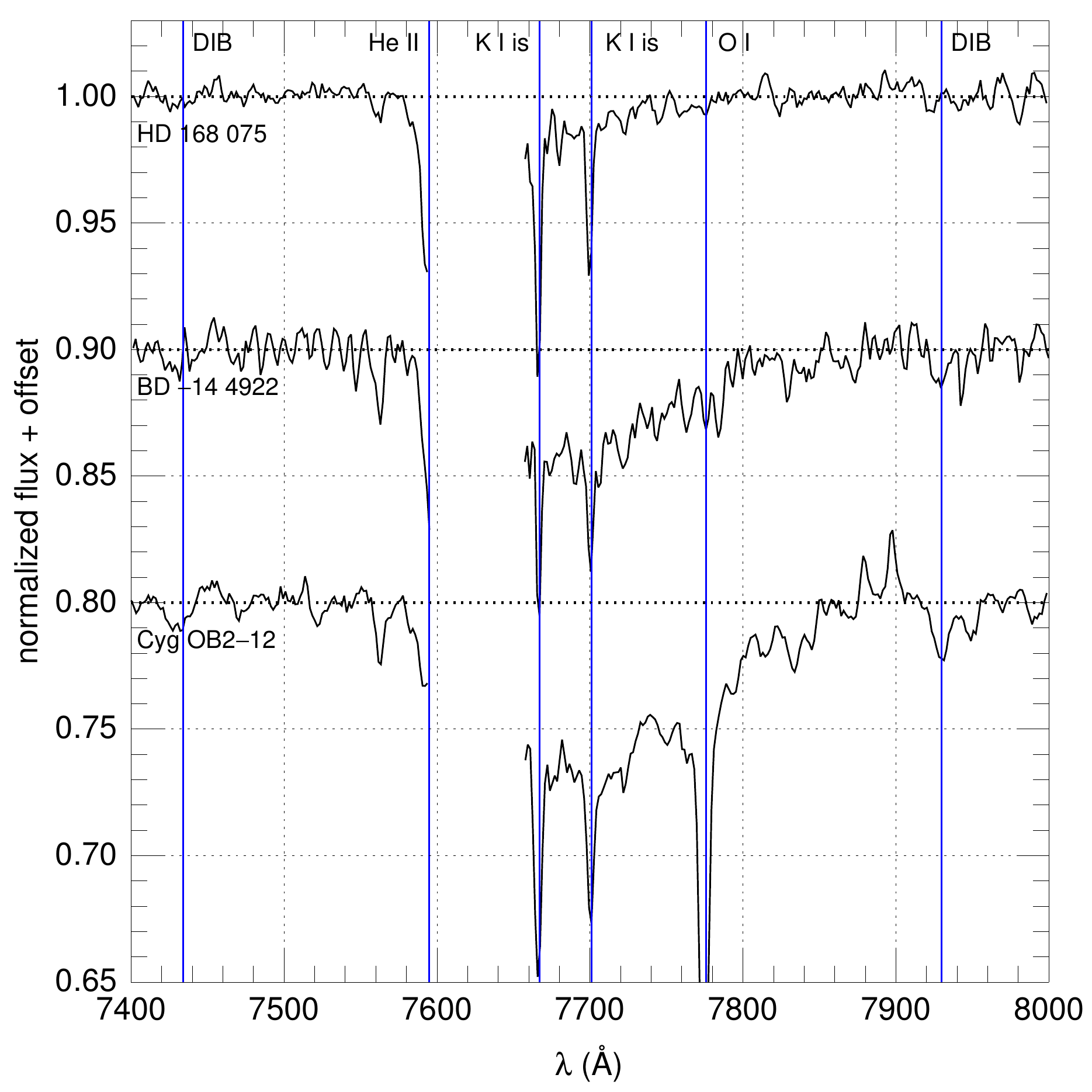}}
 \caption{Three sample LT spectra sorted by \EWb. The main stellar and ISM absorption lines have been left in place marked with
          blue vertical lines but the wavelength region with the strongest O$_2$ telluric absorption has been eliminated.}
\label{spectra_LT}
\end{figure}

\begin{figure}
\centerline{\includegraphics[width=\linewidth]{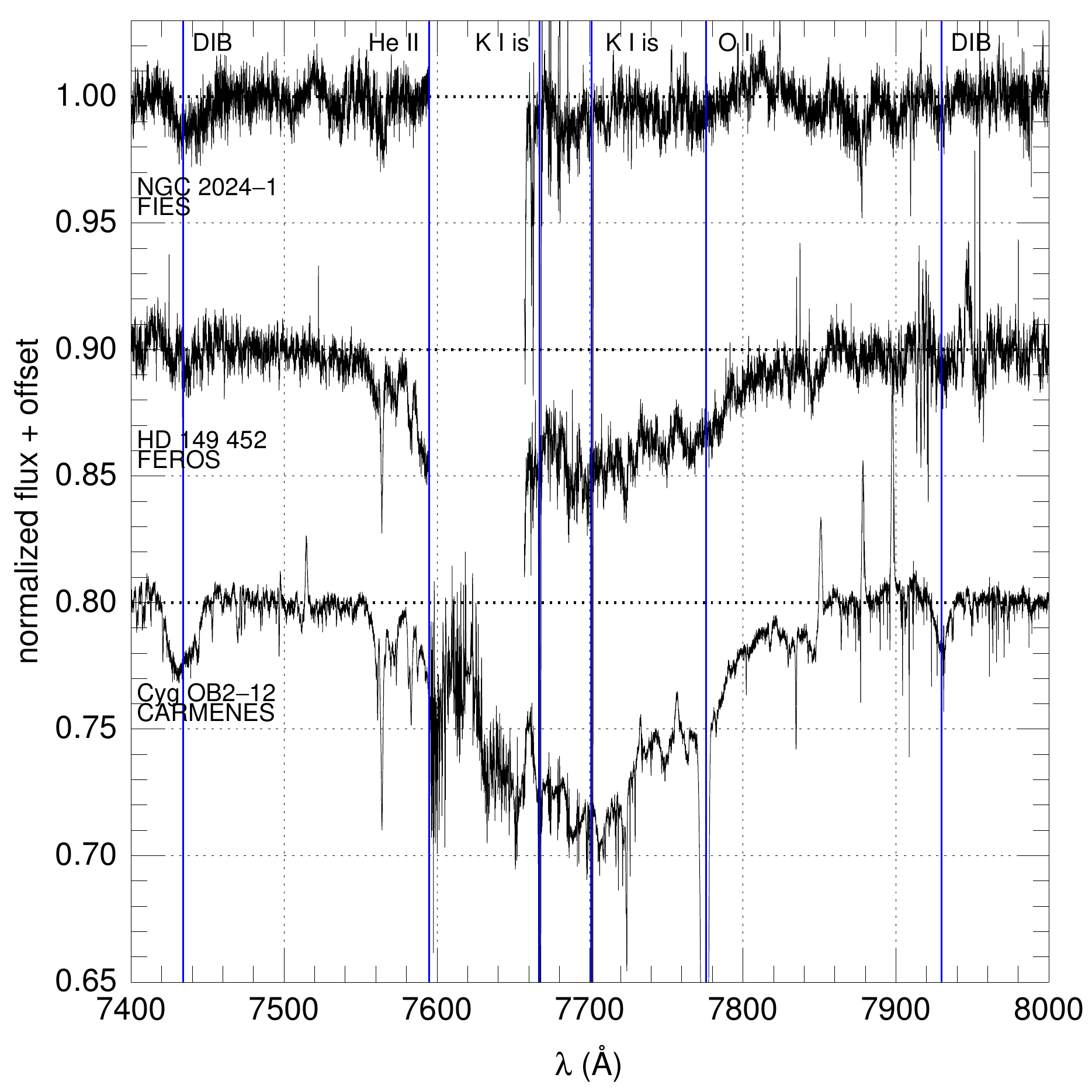}}
 \caption{Three sample \'echelle spectra sorted by \EWb. The main stellar and ISM absorption lines have been left in place marked with
          blue vertical lines but the wavelength region with the strongest O$_2$ telluric absorption has been eliminated for the first two
          spectra. For the third spectrum we have maintained that region due to two exceptional circumstances: CARMENES is a highly stable 
          spectrograph and the shown spectrum is a combination of 73 different epochs. As the telluric lines shift in wavelength
          with respect to the heliocentric frame of reference at different times of the year, we can combine the different epochs to minimize 
          their effect even in a region with strong telluric absorption such as the O$_2$ A band.}
\label{spectra_HR}
\end{figure}

For the rest of the spectrographs we first selected the sample in common between each one of them and the STIS O4--B1 sample and calculated the average and dispersion between the raw 
\EWb\ for each spectrograph pair. 
We used the average to correct the systematic bias and the dispersion to estimate the random uncertainty of the measurements of each spectrograph (Table~\ref{spectrographs}). We 
then compared the results from each spectrograph to verify the results. Table~\ref{spectrographs} indicates that the lowest uncertainties are those of LT (FRODOspec being a single-order
spectrograph does not require order stitching) and, especially, CARMENES (a spectrograph of exquisite stability designed for planet hunting with extreme precision). Once the \EWb\ from
different sources are obtained in this way, the values for objects with multiple sources are combined using their weighted means (Table~\ref{bandtable}). Sample ground-based data are 
shown in Figs.~\ref{spectra_LT}~and~\ref{spectra_HR}.

\begin{table}
\addtolength{\tabcolsep}{-3pt}
\caption{Results for the 7700~\AA\ absorption band for OBA stars.}
\centerline{
\begin{tabular}{lcccr@{$\pm$}l}
star                   & code  & RA           & dec            & \mcii{EW}    \\
                       &       & (J2000)      & (J2000)        & \mcii{(\AA)} \\
\hline
$\delta$ Ori Aa,Ab     & CFH   & 05:32:00.398 & $-$00:17:56.69 &  0.8&0.5     \\
$\mu$ Col              & SNH   & 05:45:59.895 & $-$32:18:23.18 &  0.0&0.4     \\
$\lambda$ Lep          & SH    & 05:19:34.525 & $-$13:10:36.43 &  0.1&0.4     \\
15 Mon Aa,Ab           & LCH   & 06:40:58.656 & $+$09:53:44.71 &  0.3&0.4     \\
$\rho$ Leo A,B         & SLH   & 10:32:48.671 & $+$09:18:23.71 &  0.8&0.3     \\
$\theta^{1}$ Ori Ca,Cb & LCFNH & 05:35:16.463 & $-$05:23:23.18 &  1.1&0.4     \\
$\sigma$ Ori Aa,Ab,B   & LCF   & 05:38:44.765 & $-$02:36:00.25 &  1.8&0.4     \\
HD 93\,028             & SF    & 10:43:15.340 & $-$60:12:04.21 &  2.3&0.4     \\
$\theta^{2}$ Ori A     & CFNH  & 05:35:22.900 & $-$05:24:57.80 &  0.8&0.5     \\
NU Ori                 & SH    & 05:35:31.365 & $-$05:16:02.60 &  4.4&0.4     \\
$\sigma$ Sco Aa,Ab     & LFH   & 16:21:11.313 & $-$25:35:34.09 &  4.5&0.5     \\
$\lambda$ Ori A        & CFH   & 05:35:08.277 & $+$09:56:02.96 &  1.2&0.5     \\
HD 164\,402            & SLN   & 18:01:54.380 & $-$22:46:49.06 &  2.8&0.3     \\
HD 46\,966 Aa,Ab       & SH    & 06:36:25.887 & $+$06:04:59.47 &  0.6&0.4     \\
HD 93\,205             & F     & 10:44:33.740 & $-$59:44:15.46 &  0.0&1.8     \\
9 Sgr A,B              & SFH   & 18:03:52.446 & $-$24:21:38.64 &  2.3&0.4     \\
$\alpha$ Cam           & SC    & 04:54:03.011 & $+$66:20:33.58 &  1.4&0.3     \\
$\zeta$ Oph            & FNH   & 16:37:09.530 & $-$10:34:01.75 &  2.1&0.7     \\
CPD $-$59 2591         & SF    & 10:44:36.688 & $-$59:47:29.63 &  4.6&0.4     \\
HD 34\,656             & CH    & 05:20:43.080 & $+$37:26:19.23 &  2.4&0.5     \\
Herschel 36            & LFN   & 18:03:40.333 & $-$24:22:42.74 &  2.0&0.5     \\
V662 Car               & F     & 10:45:36.318 & $-$59:48:23.37 & 10.0&1.8     \\
CE Cam                 & LCH   & 03:29:54.746 & $+$58:52:43.51 &  5.7&0.4     \\
HD 93\,162             & F     & 10:44:10.389 & $-$59:43:11.09 &  0.1&1.8     \\
HD 192\,639            & SLNH  & 20:14:30.429 & $+$37:21:13.83 &  5.2&0.3     \\
HD 93\,250 A,B         & SF    & 10:44:45.027 & $-$59:33:54.67 &  2.6&0.4     \\
BD $-$16 4826          & N     & 18:21:02.231 & $-$16:01:00.94 &  8.2&2.0     \\
HD 46\,150             & SCF   & 06:31:55.519 & $+$04:56:34.27 &  4.2&0.3     \\
CPD $-$59 2626 A,B     & F     & 10:45:05.794 & $-$59:45:19.60 &  1.0&1.8     \\
HD 149\,452            & SF    & 16:37:10.514 & $-$47:07:49.85 &  9.4&0.4     \\
HD 93\,129 Aa,Ab       & F     & 10:43:57.462 & $-$59:32:51.27 &  1.1&1.8     \\
HD 93\,161 A           & F     & 10:44:08.840 & $-$59:34:34.49 &  5.4&1.8     \\
HDE 326\,329           & F     & 16:54:14.106 & $-$41:50:08.48 &  5.1&1.8     \\
HD 48\,279 A           & SLF   & 06:42:40.548 & $+$01:42:58.23 &  3.2&0.3     \\
HDE 322\,417           & F     & 16:58:55.392 & $-$40:14:33.34 &  3.1&1.8     \\
AE Aur                 & SCH   & 05:16:18.149 & $+$34:18:44.33 &  3.0&0.3     \\
CPD $-$47 2963 A,B     & F     & 08:57:54.620 & $-$47:44:15.71 &  3.4&1.8     \\
$\lambda$ Cep          & SLNH  & 22:11:30.584 & $+$59:24:52.25 &  2.0&0.3     \\
HD 199\,216            & SLH   & 20:53:52.404 & $+$49:32:00.33 &  2.8&0.3     \\
NGC 2024-1             & CN    & 05:41:37.853 & $-$01:54:36.48 &  2.0&0.6     \\
ALS 15\,210            & F     & 10:44:13.199 & $-$59:43:10.33 &  0.0&1.8     \\
HD 46\,223             & CH    & 06:32:09.306 & $+$04:49:24.73 &  4.0&0.5     \\
HDE 319\,702           & F     & 17:20:50.610 & $-$35:51:45.97 &  3.7&1.8     \\
ALS 19\,613 A          & CN    & 18:20:29.902 & $-$16:10:44.33 & 15.8&0.6     \\
Cyg OB2-11             & N     & 20:34:08.513 & $+$41:36:59.42 & 14.9&2.0     \\
HDE 319\,703 A         & F     & 17:19:46.156 & $-$36:05:52.34 &  9.1&1.8     \\
HT Sge                 & LCFH  & 19:27:26.565 & $+$18:17:45.19 & 10.3&0.4     \\
ALS 2063               & F     & 10:58:45.475 & $-$61:10:43.01 &  6.8&1.8     \\
Cyg OB2-1 A            & N     & 20:31:10.543 & $+$41:31:53.47 & 14.8&2.0     \\
Cyg OB2-15             & N     & 20:32:27.666 & $+$41:26:22.08 & 12.5&2.0     \\
ALS 4962               & FN    & 18:21:46.166 & $-$21:06:04.42 &  7.4&1.3     \\
Cyg OB2-20             & N     & 20:31:49.665 & $+$41:28:26.51 & 14.9&2.0     \\
BD $+$36 4063          & N     & 20:25:40.608 & $+$37:22:27.07 &  5.6&2.0     \\
ALS 19\,693            & F     & 17:25:29.167 & $-$34:25:15.74 &  6.8&1.8     \\
BD $-$13 4930          & SLF   & 18:18:52.674 & $-$13:49:42.60 &  5.6&0.3     \\
HD 156\,738 A,B        & F     & 17:20:52.656 & $-$36:04:20.54 &  6.2&1.8     \\
CPD $-$49 2322         & F     & 09:15:52.787 & $-$50:00:43.82 &  5.0&1.8     \\
BD $-$14 5014          & N     & 18:22:22.310 & $-$14:37:08.46 &  6.8&2.0     \\
W 40 OS 1a             & C     & 18:31:27.837 & $-$02:05:23.66 &  3.3&0.6     \\
NGC 1624-2             & N     & 04:40:37.248 & $+$50:27:40.96 &  7.6&2.0     \\
\hline
\multicolumn{4}{l}{Codes: S,STIS; L,FRODOspec; C, CARMENES;} \\
\multicolumn{4}{l}{\phantom{Codes: }F, FEROS; N, FIES; H, HERMES.} \\
\end{tabular}
$\;\;$}
\label{bandtable}
\addtolength{\tabcolsep}{3pt}
\end{table}

\addtocounter{table}{-1}

\begin{table}
\addtolength{\tabcolsep}{-3pt}
\caption{(continued).}
\centerline{
\begin{tabular}{lcccr@{$\pm$}l}
star                   & code  & RA           & dec            & \mcii{EW}    \\
                       &       & (J2000)      & (J2000)        & \mcii{(\AA)} \\
\hline
HD 207\,198            & SLCH  & 21:44:53.278 & $+$62:27:38.04 &  2.0&0.3     \\
BD $-$12 4979          & CF    & 18:18:03.112 & $-$12:14:34.28 & 10.0&0.6     \\
Cyg OB2-B18            & H     & 20:34:57.858 & $+$41:43:54.22 & 13.5&0.9     \\
BD $+$60 513           & SN    & 02:34:02.530 & $+$61:23:10.87 &  5.7&0.4     \\
HD 217\,086            & SLNH  & 22:56:47.194 & $+$62:43:37.60 &  5.2&0.3     \\
Cyg OB2-4 B            & LN    & 20:32:13.117 & $+$41:27:24.25 & 13.8&0.6     \\
BD $-$14 4922          & LN    & 18:11:58.104 & $-$14:56:08.97 &  9.7&0.6     \\
HDE 319\,703 Ba,Bb     & F     & 17:19:45.050 & $-$36:05:47.00 & 11.5&1.8     \\
Tyc 8978-04440-1       & F     & 12:11:18.531 & $-$62:29:43.53 &  5.3&1.8     \\
HD 168\,076 A,B        & F     & 18:18:36.421 & $-$13:48:02.38 &  3.7&1.8     \\
HD 168\,112 A,B        & LFN   & 18:18:40.868 & $-$12:06:23.39 &  6.8&0.5     \\
LS I $+$61 303         & N     & 02:40:31.667 & $+$61:13:45.53 &  5.0&2.0     \\
LS III $+$46 12        & NH    & 20:35:18.566 & $+$46:50:02.90 &  4.9&0.8     \\
HD 194\,649 A,B        & LNH   & 20:25:22.124 & $+$40:13:01.07 &  6.4&0.5     \\
HDE 323\,110           & F     & 17:21:15.794 & $-$37:59:09.58 &  9.2&1.8     \\
BD $-$14 5040          & N     & 18:25:38.896 & $-$14:45:05.70 &  5.3&2.0     \\
MY Ser Aa,Ab           & FN    & 18:18:05.895 & $-$12:14:33.29 &  6.6&1.3     \\
Cyg OB2-8 A            & NH    & 20:33:15.078 & $+$41:18:50.51 & 10.6&0.8     \\
BD $+$61 487           & LN    & 02:50:13.632 & $+$62:05:30.81 &  8.5&0.6     \\
BD $-$11 4586          & CF    & 18:18:03.344 & $-$11:17:38.83 &  9.0&0.6     \\
Cyg OB2-5 A,B          & LNH   & 20:32:22.422 & $+$41:18:18.91 & 10.0&0.5     \\
HD 168\,075            & SLCFN & 18:18:36.043 & $-$13:47:36.46 &  4.8&0.3     \\
Tyc 7370-00460-1       & FN    & 17:18:15.396 & $-$34:00:05.94 & 12.7&1.3     \\
Cyg OB2-7              & NH    & 20:33:14.115 & $+$41:20:21.86 &  8.2&0.8     \\
V479 Sct               & FN    & 18:26:15.045 & $-$14:50:54.33 &  4.7&1.3     \\
BD $-$13 4923          & LFN   & 18:18:32.732 & $-$13:45:11.88 &  6.0&0.5     \\
Sh 2-158 1             & NH    & 23:13:34.435 & $+$61:30:14.73 & 15.0&0.8     \\
HD 166\,734            & LF    & 18:12:24.656 & $-$10:43:53.04 &  6.5&0.6     \\
BD $+$62 2078          & N     & 22:25:33.579 & $+$63:25:02.62 &  6.5&2.0     \\
LS III $+$46 11        & LNH   & 20:35:12.642 & $+$46:51:12.12 &  6.1&0.5     \\
ALS 18\,748            & F     & 17:19:04.436 & $-$38:49:04.87 & 13.0&1.8     \\
Pismis 24-1 A,B        & FH    & 17:24:43.502 & $-$34:11:56.96 &  9.6&0.8     \\
Cyg OB2-A11            & NH    & 20:32:31.543 & $+$41:14:08.21 & 14.8&0.8     \\
Cyg X-1                & CN    & 19:58:21.677 & $+$35:12:05.81 &  5.0&0.6     \\
BD $-$13 4929          & F     & 18:18:45.857 & $-$13:46:30.83 &  1.3&1.8     \\
ALS 15\,108 A,B        & H     & 20:33:23.460 & $+$41:09:13.02 & 14.8&0.9     \\
HD 15\,570             & SCH   & 02:32:49.422 & $+$61:22:42.07 &  4.6&0.3     \\
ALS 19\,307            & N     & 19:50:01.087 & $+$26:29:34.36 & 15.5&2.0     \\
Cyg OB2-3 A,B          & NH    & 20:31:37.509 & $+$41:13:21.01 & 15.5&0.8     \\
ALS 15\,133            & N     & 20:31:18.330 & $+$41:21:21.66 & 19.7&2.0     \\
Cyg OB2-73             & N     & 20:34:21.930 & $+$41:17:01.60 & 16.1&2.0     \\
NGC 3603 HST-5         & F     & 11:15:07.640 & $-$61:15:17.56 & 11.0&1.8     \\
Cyg OB2-22 A           & H     & 20:33:08.760 & $+$41:13:18.62 & 13.1&0.9     \\
Cyg OB2-B17            & N     & 20:30:27.302 & $+$41:13:25.31 & 12.5&2.0     \\
THA 35-II-42           & F     & 10:25:56.505 & $-$57:48:43.50 &  7.4&1.8     \\
ALS 15\,131            & N     & 20:33:02.922 & $+$41:17:43.13 & 16.3&2.0     \\
HD 17\,603             & CH    & 02:51:47.798 & $+$57:02:54.46 &  5.2&0.5     \\
Bajamar star           & LCNH  & 20:55:51.255 & $+$43:52:24.67 &  0.5&0.4     \\
Cyg OB2-22 C           & N     & 20:33:09.598 & $+$41:13:00.54 & 17.0&2.0     \\
BD $+$66 1674          & LNH   & 00:02:10.236 & $+$67:25:45.21 &  5.4&0.5     \\
Cyg OB2-27 A,B         & N     & 20:33:59.528 & $+$41:17:35.48 & 10.8&2.0     \\
BD $+$66 1675          & LNH   & 00:02:10.287 & $+$67:24:32.22 &  4.2&0.5     \\
Cyg OB2-12             & LCH   & 20:32:40.959 & $+$41:14:29.29 & 14.4&0.4     \\
BD $-$13 4927          & F     & 18:18:40.091 & $-$13:45:18.57 &  2.0&1.8     \\
Cyg OB2-9              & N     & 20:33:10.733 & $+$41:15:08.21 & 18.0&2.0     \\
V889 Cen               & F     & 13:26:59.834 & $-$62:01:49.34 & 10.7&1.8     \\
Tyc 4026-00424-1       & CH    & 00:02:19.027 & $+$67:25:38.55 &  3.2&0.5     \\
HDE 326\,775           & F     & 17:05:31.316 & $-$41:31:20.12 &  4.1&1.8     \\
V747 Cep               & NH    & 00:01:46.870 & $+$67:30:25.13 &  5.4&0.8     \\
KM Cas                 & N     & 02:29:30.477 & $+$61:29:44.14 &  8.4&2.0     \\
\hline
\multicolumn{4}{l}{Codes: S:STIS; L:FRODOspec; C, CARMENES;} \\
\multicolumn{4}{l}{\phantom{Codes: }F, FEROS; N, FIES; H, HERMES.} \\
\end{tabular}
$\;\;$}
\addtolength{\tabcolsep}{3pt}
\end{table}

\subsection{Measuring the ISM properties}

$\,\!$\indent To analyse the behaviour of the 7700~\AA\ band we need to compare the measurements of \EWb\ with other measurements of the ISM in the sightline. We use four types of data,
one of a qualitative type and three quantitative ones. 

The qualitative data that we use to analyse the behaviour of the 7700~\AA\ band are the possible presence of an H\,{\sc ii} region and its associated molecular gas around each 
object in a manner similar to what we did in paper~0. In particular, we will consider whether the star is located in a bright part of an H\,{\sc ii} region and 
whether the associated dust lanes are located between the star and us or not.

The first quantitative data are the EWs of the \KI{7701.093} ISM line coupled with the other quantities we have measured for the absorption doublet. As previously mentioned, 
Appendix~A gives the details of how we have measured the relevant quantities and results are given in Table~\ref{KI_W90_EWrat}.

The second quantitative data are the EW of the $Q$(2) (3,0) C$_2$ Phillips band line at 7724.221~\AA, which is an indicator of the existence of molecular gas in the sightline.
See Appendix~A and Table~\ref{C2table}.

The last quantitative data are the amount [\EBV] and type [\RV] of extinction as measured from optical/NIR photometry with the code CHORIZOS \citep{Maiz04c}, fixing the effective 
temperature and luminosity class from the spectral classification using the procedure described in paper~0. More specifically, in that paper we calculated \EBV\ and \RV\ for 
86 of the OBA stars in our sample here using 2MASS $JHK_{\rm s}$, {\it Gaia}~DR1~$G$, and several more optical photometric systems (including Johnson $UBV$) applying the family of extinction 
laws of \citet{Maizetal14a}. Here we use those results and apply a slightly modified version of the same procedure (substituting {\it Gaia}~DR1~$G$ for {\it Gaia}~DR2~$G$ and using 
only Johnson $UBV$ as the additional photometric system) to calculate \EBV\ and \RV\ for the rest of our OBA sample (Table~\ref{chorizos}) and, in that way, have homogeneous measurements 
of the amount and type of extinction for each star. As the process is fraught with potential pitfalls, we list some important details here:

\begin{table}
\addtolength{\tabcolsep}{-3pt}
\caption{Results for the new CHORIZOS runs for OBA stars.}
\centerline{
\begin{tabular}{lr@{$\pm$}lr@{$\pm$}lr@{$\pm$}lr}
star                   & \mcii{$E(4405-$}     & \mcii{\RV}  & \mcii{\AV}   & \chired \\
                       & \mcii{$\;\;\;5495)$} & \mcii{}     & \mcii{(mag)} &         \\
                       & \mcii{(mag)}         & \mcii{}     & \mcii{}      &         \\
\hline
$\lambda$ Lep          & 0.004&0.006          & 4.531&1.730 &  0.019&0.021 &  0.49   \\
$\rho$ Leo A,B         & 0.038&0.008          & 6.109&1.240 &  0.231&0.023 &  0.58   \\
HD 164\,402            & 0.189&0.016          & 4.210&0.486 &  0.796&0.032 &  0.78   \\
$\sigma$ Sco Aa,Ab     & 0.327&0.020          & 4.335&0.383 &  1.419&0.046 &  1.02   \\
NU Ori                 & 0.486&0.015          & 5.274&0.188 &  2.560&0.027 &  0.78   \\
BD $-$13 4930          & 0.519&0.014          & 3.341&0.122 &  1.733&0.026 &  0.78   \\
CE Cam                 & 0.573&0.023          & 3.546&0.222 &  2.033&0.080 &  1.48   \\
HD 199\,216            & 0.661&0.014          & 2.918&0.085 &  1.929&0.026 &  0.98   \\
BD $-$13 4929          & 0.876&0.034          & 3.703&0.159 &  3.243&0.030 &  0.53   \\
BD $+$61 487           & 0.883&0.039          & 4.167&0.197 &  3.680&0.031 &  1.32   \\
LS I $+$61 303         & 1.028&0.029          & 4.015&0.124 &  4.129&0.030 &  1.78   \\
ALS 15\,210            & 1.057&0.018          & 4.909&0.092 &  5.190&0.027 &  2.13   \\
BD $-$14 4922          & 1.088&0.038          & 3.365&0.123 &  3.662&0.031 &  2.34   \\
BD $-$13 4923          & 1.097&0.015          & 4.162&0.066 &  4.564&0.023 &  1.17   \\
KM Cas                 & 1.250&0.020          & 3.349&0.057 &  4.185&0.023 &  0.88   \\
HT Sge                 & 1.293&0.023          & 3.206&0.069 &  4.146&0.036 &  1.87   \\
V889 Cen               & 1.383&0.021          & 3.562&0.065 &  4.927&0.034 &  0.34   \\
Tyc 8978-04440-1       & 1.415&0.016          & 3.641&0.043 &  5.152&0.030 &  1.78   \\
Cyg OB2-20             & 1.417&0.034          & 3.055&0.071 &  4.327&0.020 &  3.35   \\
ALS 4962               & 1.424&0.021          & 3.231&0.053 &  4.601&0.026 &  0.89   \\
Cyg OB2-15             & 1.425&0.021          & 3.208&0.049 &  4.570&0.019 &  2.05   \\
Cyg OB2-4 B            & 1.438&0.030          & 2.859&0.061 &  4.112&0.022 &  0.43   \\
NGC 3603 HST-5         & 1.513&0.083          & 3.941&0.201 &  5.963&0.041 &  0.79   \\
ALS 19\,307            & 1.620&0.036          & 3.086&0.065 &  4.999&0.023 &  3.60   \\
NGC 2024-1             & 1.752&0.033          & 4.716&0.086 &  8.263&0.030 &  0.73   \\
Cyg OB2-1 A            & 1.772&0.056          & 3.110&0.087 &  5.510&0.030 &  5.78   \\
Cyg OB2-3 A,B          & 1.857&0.029          & 3.350&0.051 &  6.220&0.034 &  2.53   \\
ALS 15\,131            & 2.272&0.028          & 3.062&0.034 &  6.957&0.024 &  3.83   \\
ALS 15\,133            & 2.309&0.026          & 3.059&0.031 &  7.062&0.023 &  4.22   \\
W 40 OS 1a             & 2.440&0.111          & 4.469&0.183 & 10.903&0.071 &  2.79   \\
Cyg OB2-B18            & 2.599&0.073          & 3.178&0.076 &  8.262&0.047 &  1.30   \\
ALS 19\,613 A          & 2.736&0.027          & 3.849&0.035 & 10.532&0.028 &  6.70   \\
Cyg OB2-12             & 3.486&0.056          & 3.080&0.090 & 10.739&0.204 &  2.90   \\
Bajamar star           & 3.522&0.071          & 2.957&0.047 & 10.413&0.051 & 11.85   \\
\hline
\end{tabular}
$\;\;$}
\label{chorizos}
\addtolength{\tabcolsep}{3pt}
\end{table}

\begin{itemize}
 \item 2MASS photometry is highly uniform and has well-defined zero points \citep{MaizPant18}. Still, one has to consider the cases where it is unclear how many visual components 
       are included and correct for the effect, if needed. Also, very bright sources are saturated (leading to large uncertainties) and should be substituted by an alternative if 
       possible (e.g. \citealt{Duca02}). 
 \item {\it Gaia} photometry is extremely uniform but one should differentiate between the DR1 calibration \citep{Maiz17a} in the paper~0 results and the DR2 calibration
       \citep{MaizWeil18} for the new stars, as each one has its own passbands, zero points, and corrections. For some objects we have summed the fluxes from two nearby sources for 
       consistency with the $UBVJHK_{\rm s}$ magnitudes. We do not use the $G_{\rm BP}$ and $G_{\rm RP}$ bands here, as their measurements are problematic for a number of our 
       stars given the existence of multiple sources within the aperture used to obtain their photometry.
 \item Johnson $UBV$ magnitudes are the most heterogeneous ones. Whenever possible, we use data from large databases or surveys such as \citet{Mermetal97} or \citet{Hendetal15}.
       For the photometric calibration, see \citet{Maiz06a,Maiz07a}.
 \item As input SEDs we use the \citet{Maiz13a} effective temperature-luminosity class grid with Milky Way metallicity. The optical SEDs are from the TLUSTY OSTAR2002 \citep{LanzHube03}, 
       TLUSTY BSTAR2006 \citep{LanzHube07}, and \citet{Munaetal05} grids in decreasing order of effective temperature of the stars in our OBA sample. At longer wavelengths the Munari or 
       Kurucz models are used as the TLUSTY grid does not yield the correct colours (this was independently discovered by two of the authors here, J.M.A. and R.C.B., see
       \citealt{Bohletal17}).
\end{itemize}

For the two red giants we apply in principle the same idea as for the OBA stars and use CHORIZOS to calculate their extinction parameters. However, there is a problem: their 
{\it Gaia}~DR2~$G$ and 2MASS $JHK_{\rm s}$ magnitudes are of high quality but the available Johnson $BV$ data is poor and Johnson $U$ is just non-existent. We solved that problem by doing 
synthetic photometry for $BV$ using the STIS spectra themselves (in the wavelength range of Johnson $U$ the S/N is too low to provide a precise measurement) and combining it with the 
previously mentioned values for $GJHK_{\rm s}$. In that way and assuming 3800~K giant SEDs, for the two red giants we obtain values for \EBV\ of 2.974$\pm$0.029~mag and 2.704$\pm$0.029~mag, 
for \RV\ of 2.729$\pm$0.026 and 2.797$\pm$0.028, and for \AV\ of 8.117$\pm$0.028~mag and 7.562$\pm$0.026~mag. That is, both red giants are indeed heavily extinguished with a similar and low
\RV\ and with the first one (2MASS~J16140974$-$5147147) being more extinguished than the second one by $\sim$10\% in \EBV. The \chired\ in both cases are high so the uncertainties are
likely underestimated, something that will be discussed in the next section.

\section{Results and analysis}   

\subsection{The \EBV-\RV\ plane}

\begin{figure}
\centerline{\includegraphics[width=\linewidth]{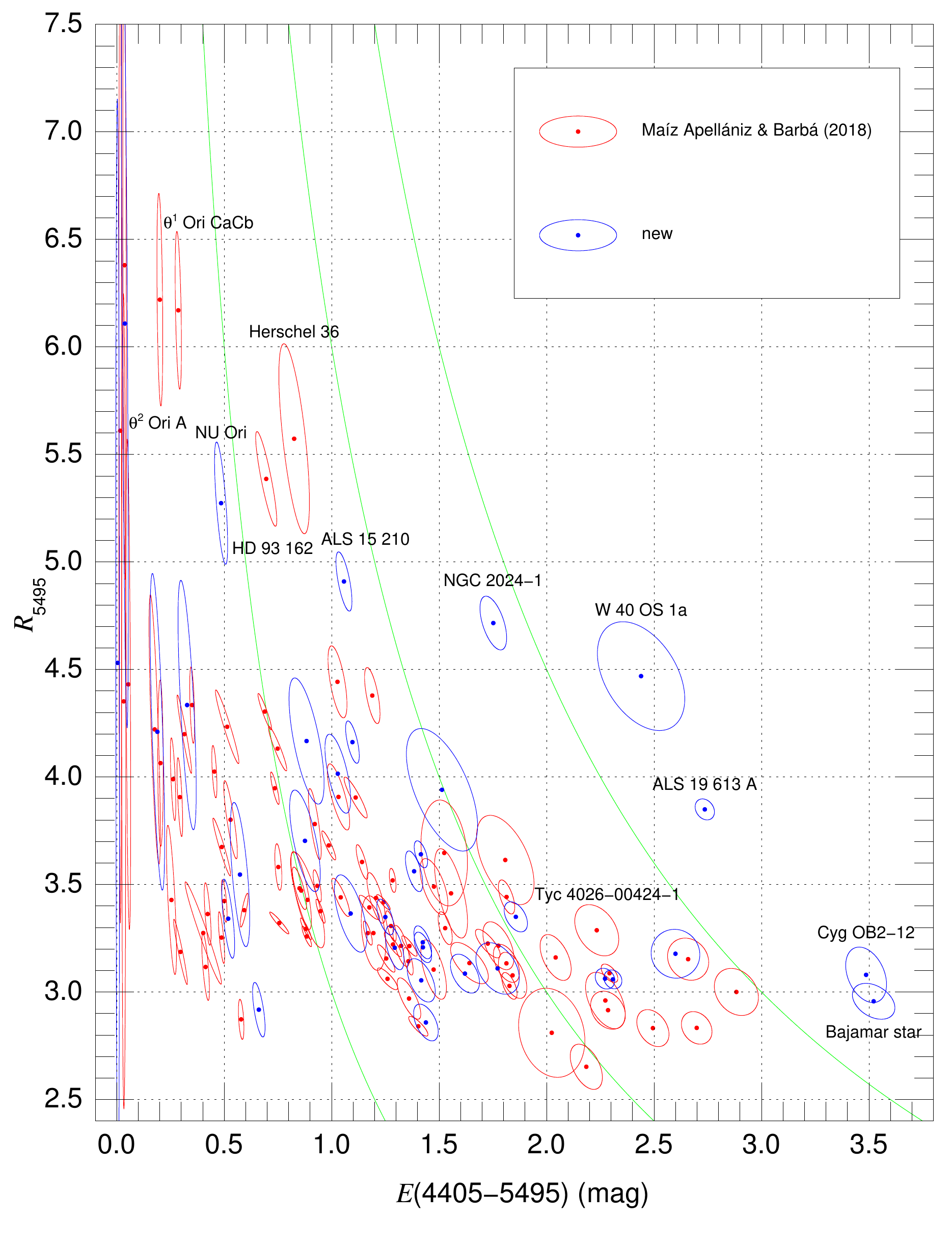}}
 \caption{\RV\ as a function of \EBV\ for the sample of 120 OBA stars in this paper. The colour coding indicates whether the data are from paper~0 or new. Some targets 
          discussed in the text are labelled. Green lines correspond (from left to right) to the values of \AV\ of 3,~6,~and~9~mag.}
\label{ebv_rv}
\end{figure}

$\,\!$\indent Before we analyse our results for the 7700~\AA\ absorption band, we start with a discussion on the general optical-NIR extinction of our sample, which is relevant to what
will be studied later. In subsection~3.3 of paper~0 we analysed the distribution of extinguished OB stars in the \EBV-\RV\ plane. Here we follow up on those results with the new values
calculated for OBA stars, which are given in Table.~\ref{chorizos}, along with those for the sample in common between the two papers. Fig.~\ref{ebv_rv} is the equivalent to 
Fig.~9 in paper~0 but with a different choice of representation. Rather than plotting the error bars for \EBV\ and \RV\ for each star we plot now the uncertainty ellipse, including 
the correlation between the two quantities as derived by CHORIZOS. As noted in Appendix~C of paper~0, the values of \EBV\ and \RV\ derived from optical+IR photometry are usually
anticorrelated and that is clearly seen in Fig.~\ref{ebv_rv}. As the ellipses are elongated in a direction close to the one defined locally by the line of constant \AV, the result is
that the relative uncertainty in \AV\ is usually less than that of \EBV\ or \RV. In other words, the CHORIZOS calculation constraints \AV\ better than either \EBV\ or \RV.

\begin{figure*}
\centerline{\includegraphics[width=\linewidth]{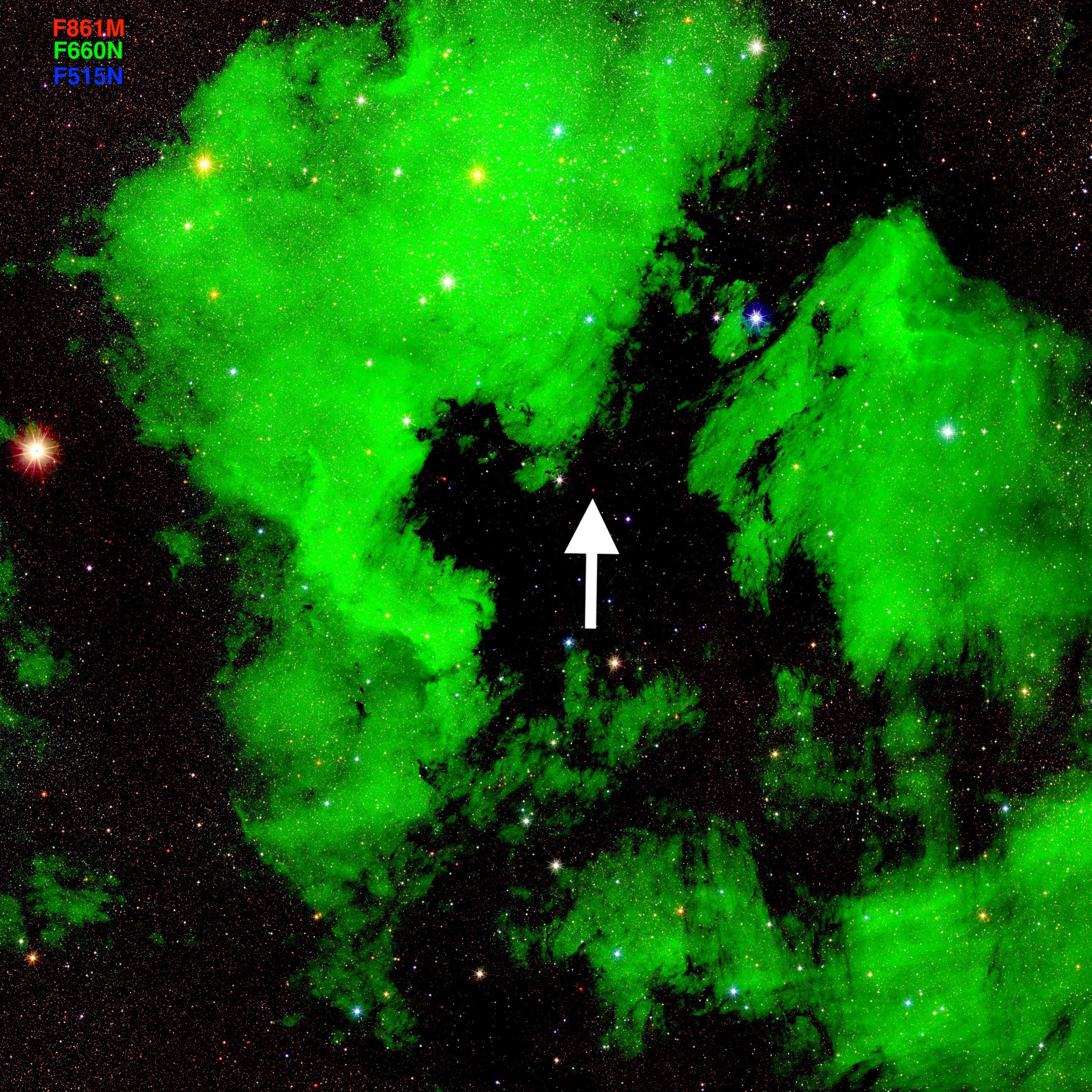}}
 \caption{The North America and Pelican nebulae as imaged by the GALANTE survey \citep{Maizetal19c,LorGetal19}. The Bajamar star, responsible for most of the ionizing photons, is the 
          red object indicated by the arrow, hidden behind the molecular cloud that creates the Atlantic Ocean and Gulf of Mexico and that also obscures the unseen nebulosity and 
          background stars. Most of the stars in the bright part of the nebulae have little extinction. The three-filter combination is composed of a Calcium triplet band (red), 
          H$\alpha$ (green), and narrow-$V$ (blue). The field size is 3.2$^{\rm o}\times$3.2$^{\rm o}$ (40~pc~$\times$~40~pc for a distance of 714~pc, Ma\'{\i}z Apell\'aniz et al. 
          submitted to A\&A) and is composed of several GALANTE fields, whose detector has a size of 1.4$^{\rm o}\times$1.4$^{\rm o}$. North is up and East is left.}
\label{North_America_Pelican}
\end{figure*}

\begin{figure*}
\centerline{\includegraphics[width=0.33\linewidth]{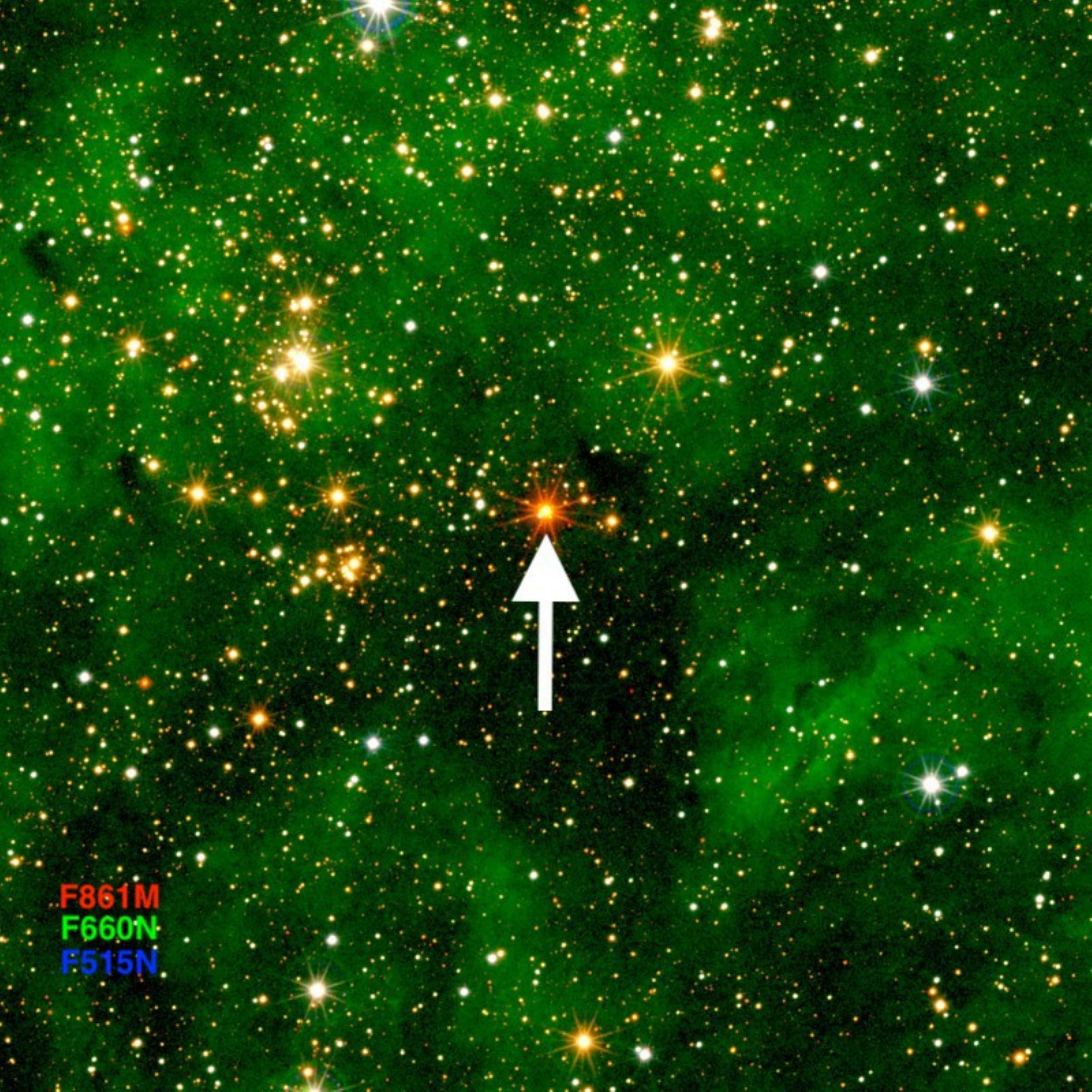} \
            \includegraphics[width=0.33\linewidth]{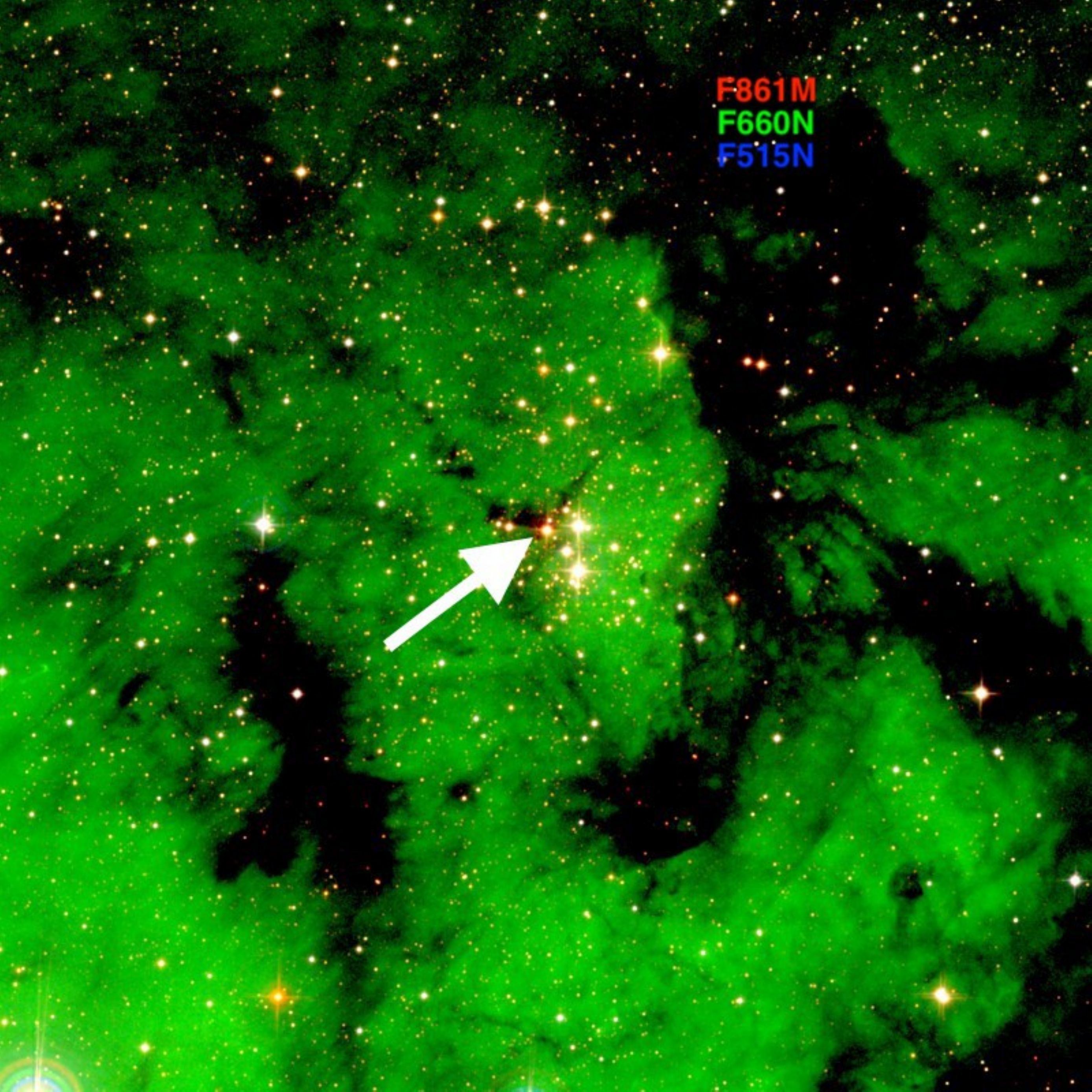} \
            \includegraphics[width=0.33\linewidth]{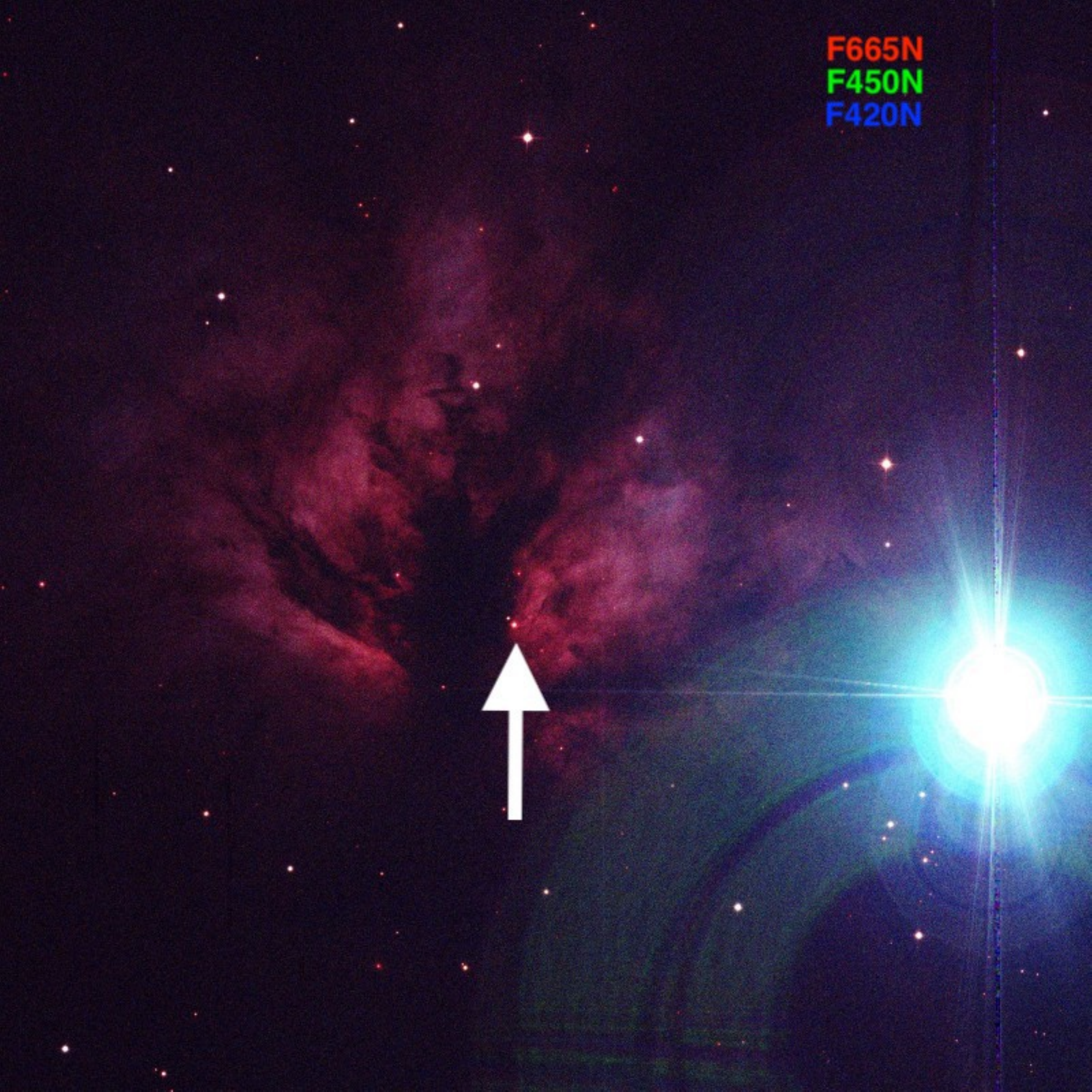}}
 \caption{Three objects in our sample imaged by the GALANTE survey \citep{Maizetal19c,LorGetal19}. [left] Cyg~OB2-12 in Cyg~OB2, where other stars in our sample are also located.
          Cyg~OB2-12 has some additional extinction with respect to the rest of the association, visible in the image as a dimming of the pervasive but diffuse nebular emission and
          by the distinct redder colour of the star. [center] Tyc~4026-00424-1 in Berkeley~59, which also contains other stars in our sample. Tyc~4026-00424-1 sits behind a dust lane
          and is redder than the surrounding OB stars. [right] NGC~2024-1 in the Flame nebula is in the bright part of the H\,{\sc ii} region but adjacent to the dust lane that 
          completely blocks its optical emission towards the East. The extremely bright source is $\zeta$~Ori, which contains two O stars and one of B type \citep{Hummetal13a,Maizetal18a}
          and is actually at a similar distance as the Flame nebula \citep{Zarietal19}, which demonstrates the effect of differential extinction. The three fields are 
          29\arcmin$\times$29\arcmin, which corresponds to linear sizes of 14.5~pc, 9.4~pc, and 3.4~pc, respectively. North is up and East is left. The first two panels use the same 
          filter combination as Fig.~\ref{North_America_Pelican} while the third one is a combination of H$\alpha$ continuum (red), continuum between H$\gamma$ and H$\beta$ (green) and 
          continuum between H$\delta$ and H$\gamma$ (blue). The nebulosity seen originating in NGC~2024 corresponds to a mixture of nebular continuum and some weak lines such as
          \HeI{6680}. Arrows indicate the object in each panel.}
\label{three_fields}
\end{figure*}

There are two main novelties in Fig.~\ref{ebv_rv} with respect to paper~0. The first one is the presence of two stars at higher \EBV\ (Bajamar~star, Fig.~\ref{North_America_Pelican}, 
and Cyg~OB2-12, Fig.~\ref{three_fields}) than any of the ones there. Both objects follow the trend discussed in paper~0 and also followed by the majority of the stars here, for which the 
average \RV\ per \EBV\ bin decreases as \EBV\ increases, with the highest bins having averages close to \RV~=~3.0. On the other hand, there are three new stars that break that trend 
(NGC~2024-1, Fig.~\ref{three_fields}; W~40~OS~1a; and ALS~19\,613~A) and appear in the previously empty region with $\EBV > 1.7$~mag and $\RV > 3.8$ i.e. they are objects with simultaneously 
large values of \EBV\ and \RV. What do these three objects have in common? They are all in the nebular-bright part of H\,{\sc ii} regions (NGC~2024, W~40, and M17, respectively) but with 
a line of sight that passes close to the molecular cloud. Indeed, other objects located towards the left and upwards from them in the plot (ALS~15\,210, HD~93\,162, and Herschel~36, the first 
two in the Carina nebula and the third in M8) share the same characteristic. Continuing in that direction in Fig.~\ref{ebv_rv} towards lower values of \EBV\ we find three of the stars in the 
Orion nebula ($\theta^1$~Ori~Ca,Cb and $\theta^2$~Ori~A) and the nearby M43 (NU~Ori). These nine stars that depart from the general \EBV-\RV\ trend will be referred in the next 
subsection as the high-\RV\ subsample. This reinforces our conclusion from paper~0: regions with high levels of UV radiation have large values of \RV. The novelty here is that 
{\bf we have apparently found the condition required for large amounts of high-\RV\ dust to exist: there must be a nearby molecular cloud to act as a source.} As we proposed in 
\citealt{Maiz15a} and in paper~0, the grains that were originally in the protected molecular environment are exposed to the intense UV radiation from the nearby O star(s) and the small 
grains (mostly PAHs, \citealt{Tiel08}) are selectively destroyed, leaving behind the large grains responsible for the high values of \RV.

Objects with very high values of \AV\ in Table~\ref{chorizos} have large values of \chired. This is likely a sign that for \AV$\sim$10~mag the \citet{Maizetal14a} family of extinction laws 
is starting to lose its accuracy, something that was expected, and that signals the need for an improved family of optical extinction laws, an issue that we will tackle in future papers of 
this series. For our immediate needs, we can consider that the uncertainties in the extinction parameters in Table~\ref{chorizos} are likely underestimated by a factor of 2--3 for the four 
stars with \AV$\sim$10~mag.

%\begin{figure}
%\centerline{\includegraphics[width=\linewidth]{chi2stats.pdf}}
% \caption{\chired\ statistics of the CHORIZOS fits for the sample of 120 OBA stars in this paper divided in two blocks of low and high extinction, respectively. Compare with
%          Fig.~4 of paper~0.}
%\label{chi2stats}
%\end{figure}

\subsection{The 7700~\AA\ absorption band}

\begin{figure}
\centerline{\includegraphics[width=\linewidth]{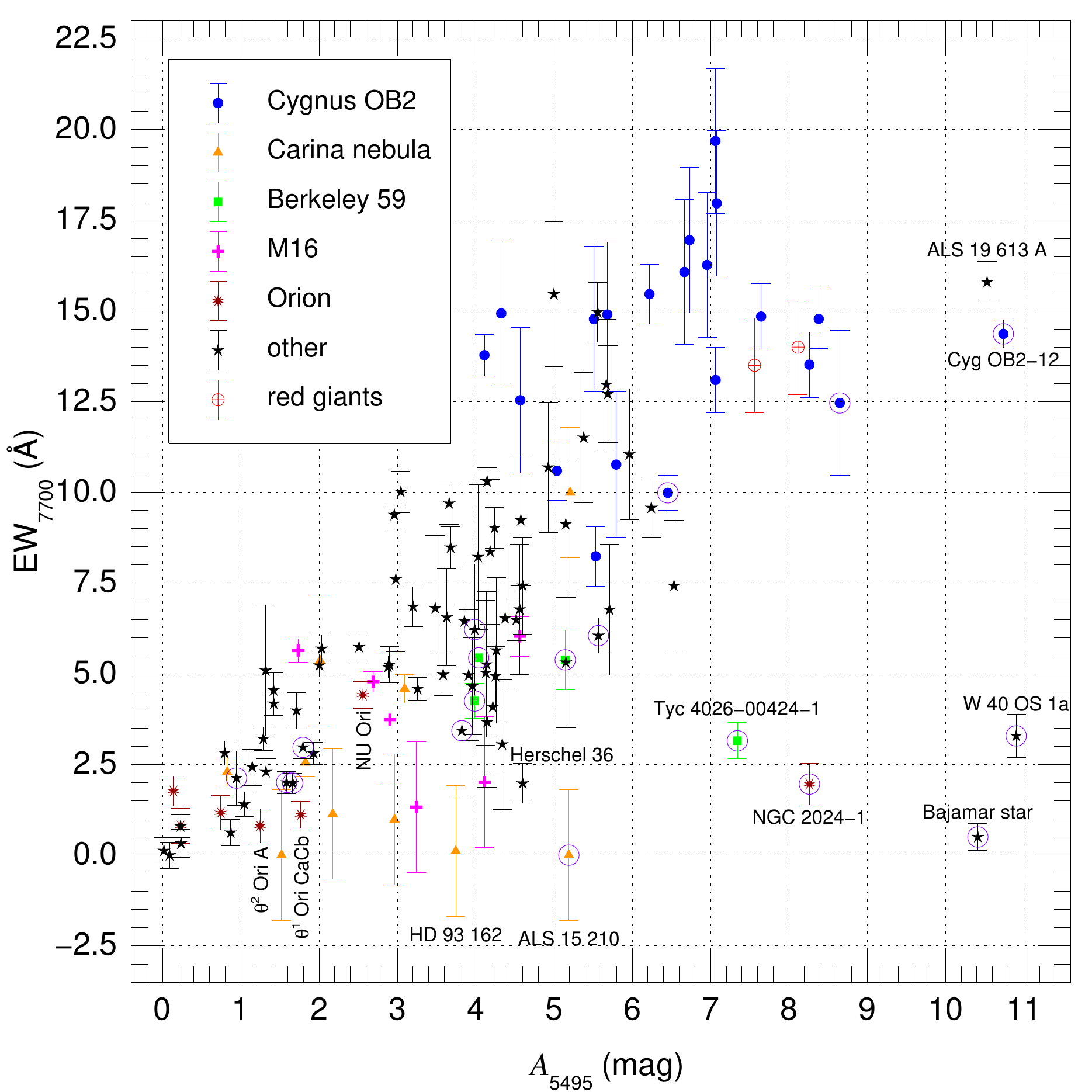}}
 \caption{\EWb\ as a function of \AV\ for the full sample in this paper. The colour coding indicates specific regions in the sky and the two red giants. In addition, objects with a 
          purple circle around them have C$_2$ in absorption detected in Table~\ref{C2table}. Orion refers to the whole
          region (e.g. \citealt{Zarietal19}), not to the Orion nebula alone. Some targets discussed in the text are labelled.}
\label{av_ew7700}
\end{figure}

\begin{figure}
\centerline{\includegraphics[width=\linewidth]{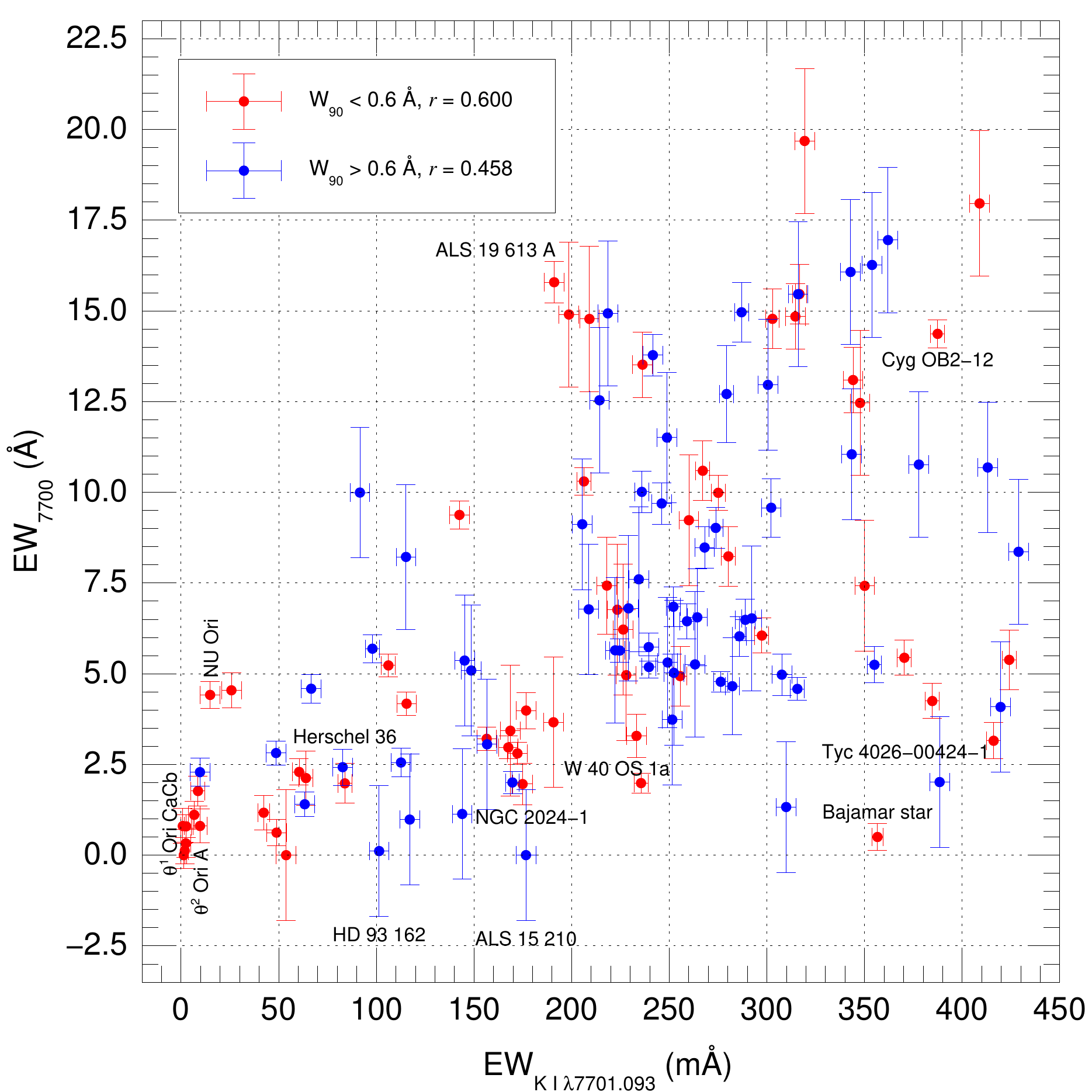}}
 \caption{\EWb\ as a function of \EWKIr\ for the sample of 120 OBA stars in this paper. Symbols are colour-coded by W$_{90}$ for \KI{7701.093}. Pearson correlation 
          coefficients for the two subsets are given in the legend. Some targets discussed in the text are labelled.}
\label{KIr_ew7700}
\end{figure}

$\,\!$\indent We start our analysis of the 7700~\AA\ absorption band by plotting its EW as a function of \AV\ (Fig.~\ref{av_ew7700}) and of \EWKIr\ (Fig.~\ref{KIr_ew7700}), 
two different measurements of the amount of intervening material in the ISM along the line of sight. In both cases there is a significant correlation but it is readily apparent that it is 
stronger for \AV\ than for \EWKIr. In some stars the different saturation levels of the line for different ISM kinematics may be at play for \EWKIr\ (see Appendix A) but this effect 
cannot be the whole story, as even for relatively narrow lines (W$_{90}<$~0.6~\AA) the Pearson correlation coefficient is just 0.601. On the other hand, the Pearson correlation 
coefficient between \AV\ and \EWb\ is 0.633 for the sample of 120 OBA stars and if we exclude the 12 named stars in Fig.~\ref{av_ew7700} it goes up to 0.814. The twelve excluded stars 
are the nine that depart from the general trend in Fig.~\ref{ebv_rv}, the two objects with the highest \EBV\ there (see previous subsection), and Tyc~4026-00424-1 (below we describe their
special circumstances). This is a first sign that the 7700~\AA\ absorption band is possibly related to the dust that extinguishes the stellar continuum but the existence of 
some outliers indicates that there must be at least one difference between its carrier and dust particles in general. 

Something else we can notice in Fig.~\ref{av_ew7700} is that the two red giants appear to follow the same general trend as the OBA stars, a sign that the carrier appears to be in the
general ISM and not in clouds specifically associated with OBA stars. Of course, two objects are a small sample that needs to be increased to confirm whether that hypothesis is correct
(we have GO~15\,816 observations pending so we should confirm this in the near future). Another, perhaps more important, effect can also be seen in Fig.~\ref{av_ew7700}, where we have
colour-coded objects by their membership to a given region. With a few exceptions, what we see is that for a given region there is little variation in \EWb\ despite large changes in 
\AV, i.e. the correlation between \AV\ and \EWb\ disappears when only objects in the same region are considered. For example, targets in Cyg~OB2 (Fig.~\ref{three_fields}) have values of
\AV\ between $\sim$4~mag and $\sim$11~mag but \EWb\ is close to 15~\AA\ for most objects and the two at the extremes of the \EWb\ distribution have intermediate values of \AV. Even more 
extremely, stars in Orion span an extinction range from close to zero to more than 8~mag in \AV\ while keeping \EWb\ around 1--2~\AA\ with the single exception of the intermediate 
extinction NU~Ori, the only one with a slightly higher value. This prompts our hypothesis with respect to {\bf the carrier of the 7700~\AA\ band: it is strongly depleted in the ISM 
associated with young star-forming regions} such as the ones where OB stars are usually located. 

In order to verify that hypothesis we now turn to another one of our measured quantities, that of the (3,0) C$_2$ Phillips band lines. As shown in the Appendix (Table~\ref{C2table}), we 
have only detected them in 18 of our 120 OBA stars (15\% of the sample), indicating that most sightlines do not have significant amounts of molecular carbon and that most of the dust is 
associated with low/intermediate density regions in the ISM. However, those 18 detections include the four objects that depart more clearly from the general trend in Fig.~\ref{av_ew7700},
i.e. those with \AV~$>$~6~mag and \EWb~$<$~5~\AA: Tyc~4026-00424-1, NGC~2024-1, Bajamar star, W~40~OS~1a. Furthermore, W~40~OS~1a is the target with the highest measured C$_2$ line 
intensity in our sample and with one of the two highest departures from the general trend in Fig.~\ref{av_ew7700} and all four of them are among the highest EWs in Table~\ref{C2table}. 
In general, the 18 detections, shown with purple circles in
Fig.~\ref{av_ew7700}, are located towards the right with respect to the general trend in that plot. Therefore, there is a clear correlation between the departure
of \EWb\ from its expected value for a given amount of extinction and the EW for the $Q$(2) (3,0) C$_2$ Phillips band line. This allows us to refine the hypothesis about {\bf the carrier 
of the 7700~\AA\ band: it is strongly depleted in the molecular clouds rich in C$_2$ in young star-forming regions}.

Let us analyse the circumstances of the four stars mentioned in the previous paragraph. They all share two characteristics: (a) they are relatively close to the Sun compared to the sample
average (Tyc~4026-00424-1 is the most distant one, at $\sim$1.1~kpc, most targets in our sample such as those in Cyg~OB2, the Carina nebula, or M16 are beyond that) and 
(b) when nearby stars have been measured they show considerably lower extinctions, e.g. most stars in the Orion region or in the bright parts of the North America nebula have low values 
of \EBV. That combination implies that they are affected mostly by local extinction but that most of the sightline
towards each star is relatively dust-free. Indeed, that is what the GALANTE images in Figs.~\ref{North_America_Pelican}~and~\ref{three_fields} show (see \citealt{Hercetal19} for the 
similar W~40 circumstances): the four stars are either behind dust lanes (where C$_2$ is likely to reside) or close to them, which likely means that due to projection effects the 
sightline crosses both the bright nebulosity and the molecular gas. This reinforces our hypothesis regarding the carrier of the 7700~\AA\ band.

What about the other stars with strong C$_2$ absorptions? Two of them are Cyg~OB2-12 and Cyg~OB2-B17. They are the most extinguished stars from Cyg~OB2 in our
sample but their values of \EWb\ are average for the association. The most likely explanation is that there is a large extinction dispersed along the sightline (which coincides with
the Cygnus arm) that affects all of the association but that those two stars have nearby molecular gas (see Fig.~\ref{three_fields} for Cyg~OB2-12\footnote{Though it is difficult to see
in Fig.~\ref{spectra_lines}, the C$_2$ lines for Cyg~OB2-12 show at least two components with different velocities and possibly more \citep{GredMunc94}.}) that introduces the 
extra extinction and the C$_2$ lines without contributing significantly to \EWb. Also, all four stars in Berkeley~59 (the previously mentioned Tyc~4026-00424-1 but also BD~$+$66~1674, 
BD~$+$66~1675, and V747~Cep) have intense C$_2$ absorption. We already pointed out in paper~0 that this cluster had a very high extinction for its distance and that it was
likely that ``most of the extinction common to the four sightlines is coming from a molecular cloud that affects all sightlines'', a prediction that we confirm here with this
detection. An additional star with a strong C$_2$ absorption is LS~III~$+$46~11 in Berkeley~90, which was
analysed in detail in \citet{Maizetal15c}. They proposed there (Fig.~4 in that paper) that the star experiences a local additional extinction that does not affect the nearby LS~III~$+$46~12 
(where we do not detect C$_2$ absorption) and that is caused by the core of a cloud that is typical of a $\zeta$-type DIB sightline. Such clouds are shielded from UV radiation, as opposed to 
those of $\sigma$-type sightlines, which are exposed to UV radiation \citep{Kreletal97,Camietal97}. This also reinforces our hypothesis, as the values of \EWb\ for those two 
stars in Berkeley~90 are just $\sim$1~sigma apart from each other, indicating that the local additional extinction that affects LS~III~$+$46~11 produces little effect in the 7700~\AA\ 
band. 

The Berkeley~90 study by \citet{Maizetal15c} prompts us to look into the relationship between the 7700~\AA\ band and DIBs. As we previously mentioned, the band is much wider that any
known DIB but it is worth noting that close to its central wavelength \citet{JennDese94} describe a DIB with central wavelength of 7711.8~\AA\ (vacuum) and a FWHM of 33.5~\AA\ (five and a
half times narrower than our measurement). It is likely that this DIB was the central part of the 7700~\AA\ band with the rest flattened by the spectrum rectification. The more recent 
compilations of \citet{Sonnetal18} and \citet{Fanetal19} do not include any broad DIBs in the vicinity. At a more general 
level, the characteristics we have described in the previous paragraphs indicate that {\bf the behaviour of the 7700~\AA\ band is similar to that of $\sigma$-type DIBs.} That
is, it originates in a carrier present in the low/intermediate density regions of the ISM exposed to UV radiation but is depleted in the dense, shielded regions. We leave for a future
study an analysis of this relationship between $\zeta$+$\sigma$ DIBs and the 7700~\AA\ band with a large sample. Here we simply point to the behaviour of \EWb\ in the two prototype 
sightlines (from whose Bayer Greek letter receive their names): $\sigma$~Sco~Aa,Ab (a complex multiple system dominated by a B1~III star, \citealt{Morgetal53b,Greletal15}) and 
$\zeta$~Oph (a runaway late-O star, \citealt{Hoogetal01,Maizetal18b}). Both have similar values of \EBV\ but $\sigma$~Sco~Aa,Ab has an \EWb\ approximately double that of $\zeta$~Oph. 
Therefore, the current data indicate a similar behaviour for the 7700~\AA\ band and $\sigma$-type DIBs and, hence, a possible common carrier. Another line of study that we plan to pursue in 
the future is the relationship between the 7700~\AA\ band and 
the so-called C$_2$-DIBs \citep{Thoretal03,Elyaetal18}, % REFEREE
which appear to be associated with high column densities of molecular carbon.

\subsection{An ISM model}

$\,\!$\indent Here we analyse the relationship between the 7700~\AA\ band and \RV. We recall that in the previous section we identified nine stars that depart from the general \EBV-\RV\ 
trend (the high-\RV\ subsample). Eight of those stars follow a nearly-horizontal trend in Fig.~\ref{av_ew7700} and include two of the four extreme objects in the lower right corner of
that figure. This suggests a connection between the 7700~\AA\ band and dust grain size, with the carrier being depleted in regions of high \RV. The relationship is not perfect, though, as 
two stars in the lower right corner of the figure (Tyc~4026-00424-1 and Bajamar star) have normal values of \RV\ and ALS~19\,613~A has simultaneously high \RV\ and normal \EWb\ values. 
A hypothesis that fits the data describes the five objects in our sample with high extinction in \HII\ regions by dividing them in three groups: [a] stars where the sightline places them 
behind the parent molecular cloud (Tyc~4026-00424-1 and Bajamar star), [b] stars with a mixed sightline or in a transition region with high \RV\ and low \EWb\ (NGC~2024-1 and W~40~OS~1a),
and [c] stars with high \RV\ and normal \EWb\ (ALS~19\,613~A). Objects in the first group experience ``molecular-cloud extinction'' with low \RV, little or no carrier of the 7700~\AA\ 
band, and C$_2$ absorption. For those in the second group destruction of small grains has started (increasing \RV) but C$_2$ is still present and the production of the carrier of the 
7700~\AA\ band has not started. Finally, in the third group the carrier of the 7700~\AA\ band already exists and C$_2$ is likely destroyed, producing an ``\HII-region extinction'' with
high \RV, presence of the 7700~\AA-band carrier, and no molecular carbon. Away from \HII\ regions we have the general trend in \EBV-\RV\ that goes from \RV\ of 4.0-4.5 for low \EBV\
(with a high dispersion) to values of 3.0 for high \EBV. In paper~0 we proposed that this trend is an effect of the increasing density and decreasing UV radiation field as one goes from the
diffuse to the translucent ISM (e.g. Fig.~1 in \citealt{SnowMcCa06}) that dominates extinction for most Galactic sightlines. That ``typical'' extinction has no C$_2$ and the 7700~\AA\ band
is present throughout, either in the high-\RV\ environment of the Local Bubble and similar diffuse ISM regions or in the denser clouds that produce an \RV\ close to 3.0.

The model described in the previous paragraph should be tested with better data, especially with high-extinction stars in \HII\ regions. For the stars in \HII\ regions in the high-\RV\ 
subsample not mentioned there we can assign such a classification (molecular-cloud, intermediate, or \HII-region extinctions) based on their location in Fig.~\ref{av_ew7700}, with 
HD~93\,162, Herschel~36, and ALS~15\,210 (for which we indeed detect C$_2$ in absorption) in the second group; NU~Ori in the third; and $\theta^1$~Ori~Ca,Cb and $\theta^2$~Ori~A in either the
second or third. Spectra with higher S/N (the values in our current data change significantly from star to star) are required to better detect C$_2$ and observations of more stars are needed 
to test the model. However, we expect the real world to be more complicated. In a real sightline, extinction is produced not in a single phase of the ISM
but in a combination of them. While absorption structures (whether atomic, molecular, DIB, or the new band) can be superimposed on a spectrum being present in just one of the phases
(and diluted in the rest if the carrier is not present there), the value of \RV\ is a weighted average of the individual values of each phase of the ISM (see Appendix~C of 
\citet{Maizetal14a}, note this is true for families of extinction laws of the form $A_\lambda/\AV = a(\lambda) + b(\lambda)/\RV$). Therefore, as information from different phases is 
preserved (although diluted) in the absorption structures but simply combined into \RV\ to give an average, it should be possible to combine ISM phases to build two sightlines with 
identical \RV\ but different absorption spectra. \RV\ can be correlated with values derived from ISM absorption lines but the correlation can never be perfect (Fig.~5 in
\citealt{Maiz15a}). The identity of the carrier of the 7700~\AA\ band is unknown at this time but we know it is abundant in typical ISM sightlines and depleted in those with abundant 
molecular gas and that is possibly the same as that of the $\sigma$-type DIBs.

%\subsection{The detailed behaviour of the extinction law in the optical range}
%
%$\,\!$\indent 
%

\section{Summary and future work}

$\,\!$\indent In this work we report the discovery of a broad interstellar band centred around 7700~\AA\ that had remained hidden behind an O$_2$ telluric band. Its equivalent width 
correlates well with the amount of extinction but deviates from the correlation for some sightlines, particularly those that are rich in molecular carbon, where its values are lower. A 
similar depletion of its carrier in such dense environments also takes place for $\sigma$-type DIBs, pointing towards a possible common origin for both types of ISM absorption features. We 
also extend the model of paper~0 to describe the different types of extinction that take place in star-forming regions with two extreme types: a ``molecular-cloud extinction'' with low \RV, 
low \EWb, and C$_2$ absorption and an ``\HII-region extinction'' with high \RV, high \EWb, and no C$_2$ absorption, with some sightlines falling between the two extremes. Away from
star-forming regions, the 7700~\AA\ band is ubiquitous, C$_2$ is absent, and \RV\ decreases as the density of the medium increases and the UV radiation field decreases. Our results are based
on 26, 30, and 120 stars observed with low-resolution long-slit space-based STIS, intermediate-resolution IFU ground-based FRODOspec, and high-resolution \'echelle ground-based spectroscopy,
respectively. In future papers we 
plan to address the detailed behaviour of the extinction law in the optical range using spectrophotometry, connect the optical and IR extinction laws, and analyse a large sample of 
interstellar absorption lines to produce a better understanding of the relationship between extinction and the different phases of the ISM.

\section*{Acknowledgements}
$\,\!$\indent We thank 
an anonymous referee for useful comments and % REFEREE
the STScI and earthbound telescope staffs for their help in acquiring the data for this paper.
J.M.A. and C.F. acknowledge support from the Spanish Government Ministerio de Ciencia through grant PGC2018-095\,049-B-C22. 
R.H.B. acknowledges support from DIDULS Project 18\,143 and the ESAC Faculty Visitor Program.
J.A.C. acknowledges support from the Spanish Government Ministerio de Ciencia through grant AYA2016-79\,425-C3-2-P. 

\section*{Data availability} 
$\,\!$\indent This paper uses observations made with the ESA/NASA Hubble Space Telescope, obtained from the data archive at the Space Telescope Science Institute. 
STScI is operated by the Association of Universities for Research in Astronomy, Inc. under NASA contract NAS 5-26\,555. 
The ground-based spectra in this paper have been obtained with the 2.2~m Observatorio de La Silla Telescope, the 3.5~m Telescope at the Observatorio de Calar Alto (CAHA), 
%the 8.2~m Kueyen Telescope at Observatorio Paranal, 
and three telescopes at the Observatorio del Roque de los Muchachos (ORM): the 1.2~m Mercator Telescope (MT), the 2.0~m Liverpool Telescope (LT), and the 2.5~m Nordic 
Optical Telescope (NOT). Some of the MT and NOT data were obtained from the IACOB spectroscopic database \citep{SimDetal11a,SimDetal11c,SimDetal15b}.
The GALANTE images were obtained with the JAST/T80 telescope at the Observatorio Astrof{\'\i}sico de Javalambre, Teruel, Spain (owned, managed, 
and operated by the Centro de Estudios de F{\'\i}sica del Cosmos de Arag\'on). 
No reflective surface with a size of 4~m or larger was required to write this paper.
The derived data generated in this research will be shared on reasonable request to the corresponding author for those cases where no proprietary rights exist.
%------------------------------------------------------------------
\bibliographystyle{mnras} % style mnras.bst
\bibliography{general} % your references references.bib

%\vfill
%
%\eject
%
%$\,\!$
%
%\vfill
%
%\eject

%%%%%%%%%%%%%%%%%%%%%%%%%%%%%%%%%%%%%%%%%%%%%%%%%%

%%%%%%%%%%%%%%%%% APPENDICES %%%%%%%%%%%%%%%%%%%%%

\appendix
\section{ISM lines in the 7400--8000~\AA\ range}

\begin{table}
\caption{Main ISM absorption lines in the 7400--8000~\AA\ region (vacuum wavelengths). The last column indicates whether it is a new line (yes), an old one (no), or an old one
         with new measurements (meas). BIB stands for Broad Interstellar Band.}
\label{ISMlines}
\centerline{
\begin{tabular}{lr@{.}lr@{.}lc}
type          & \mcii{$\lambda_{\rm c}$} & \mcii{FWHM}    & new? \\
              & \mcii{($\AA$)}           & \mcii{($\AA$)} &      \\
\hline
DIB           & 7434&1(0.2)              &  19&6(0.2)     & meas \\
K\,{\sc i}    & 7667&021                 & \mcii{---}     & no   \\
BIB           & 7700&0(1.3)              & 179&9(3.8)     & yes  \\
K\,{\sc i}    & 7701&093                 & \mcii{---}     & no   \\
DIB           & 7708&88(0.03)            &   0&71(0.04)   & meas \\
DIB           & 7710&20(0.03)            &   0&65(0.03)   & meas \\
C$_2$ $R$(6)  & 7716&698                 & \mcii{---}     & no   \\
C$_2$ $R$(4)  & 7717&067                 & \mcii{---}     & no   \\
C$_2$ $R$(8)  & 7717&539                 & \mcii{---}     & no   \\
C$_2$ $R$(2)  & 7718&652                 & \mcii{---}     & no   \\
C$_2$ $R$(10) & 7719&594                 & \mcii{---}     & no   \\
C$_2$ $R$(0)  & 7721&454                 & \mcii{---}     & no   \\
DIB           & 7722&58(0.07)            &   3&70(0.09)   & meas \\
DIB           & 7724&03(0.01)            &   0&79(0.01)   & meas \\
C$_2$ $Q$(2)  & 7724&221                 & \mcii{---}     & no   \\
C$_2$ $Q$(4)  & 7726&345                 & \mcii{---}     & no   \\
C$_2$ $P$(2)  & 7727&945                 & \mcii{---}     & no   \\
C$_2$ $Q$(6)  & 7729&684                 & \mcii{---}     & no   \\
C$_2$ $P$(4)  & 7733&791                 & \mcii{---}     & no   \\
C$_2$ $Q$(8)  & 7734&245                 & \mcii{---}     & no   \\
C$_2$ $Q$(10) & 7740&033                 & \mcii{---}     & no   \\
C$_2$ $P$(6)  & 7740&867                 & \mcii{---}     & no   \\
C$_2$ $P$(8)  & 7749&169                 & \mcii{---}     & no   \\
DIB           & 7929&9(0.1)              &   9&8(0.3)     & meas \\
\hline
\end{tabular}
}
\end{table}

\begin{table}
\addtolength{\tabcolsep}{-3pt}
\caption{Results for the ISM absorption line K\,{\sc i}~$\lambda$7701.093.}
\centerline{
\begin{tabular}{lcr@{$\pm$}lr@{$\pm$}lcr@{$\pm$}l}
star                   & code & \mcii{EW}     & \mcii{$v$}    & W$_{90}$ & \mcii{$\frac{{\rm EW}_{7701.093}}{{\rm EW}_{7667.021}}$} \\
                       &      & \mcii{(m\AA)} & \mcii{(km/s)} & (\AA)    & \mcii{}                                                  \\
\hline
$\delta$ Ori Aa,Ab     & H    &   1&1         & \mcii{---}    & ---      & \mcii{---}  \\
$\mu$ Col              & NH   &   1&1         & \mcii{---}    & ---      & \mcii{---}  \\
$\lambda$ Lep          & H    &   2&2         & \mcii{---}    & ---      & \mcii{---}  \\
15 Mon Aa,Ab           & H    &   2&2         & \mcii{---}    & ---      & \mcii{---}  \\
$\rho$ Leo A,B         & H    &   3&3         & \mcii{---}    & ---      & \mcii{---}  \\
$\theta^{1}$ Ori Ca,Cb & FNH  &   7&3         & $+$19.5&0.3   & 0.30     & 0.620&0.398 \\
$\sigma$ Ori Aa,Ab,B   & CF   &   9&4         & $+$21.9&0.3   & 0.34     & 0.398&0.205 \\
HD 93\,028             & F    &  10&5         &  $+$4.2&0.4   & 0.62     & 0.270&0.200 \\
$\theta^{2}$ Ori A     & FH   &  10&4         & $+$19.4&0.3   & 0.38     & 0.484&0.289 \\
NU Ori                 & H    &  15&5         & $+$23.9&0.4   & 0.28     & 0.392&0.234 \\
$\sigma$ Sco Aa,Ab     & F    &  26&5         &  $-$6.6&0.4   & 0.37     & 0.631&0.314 \\
$\lambda$ Ori A        & CFH  &  42&3         & $+$24.8&0.3   & 0.28     & 0.477&0.043 \\
HD 164\,402            & N    &  49&5         &  $-$9.5&0.4   & 0.70     & 0.557&0.132 \\
HD 46\,966 Aa,Ab       & H    &  49&5         & $+$18.3&0.4   & 0.50     & 0.522&0.117 \\
HD 93\,205             & F    &  54&5         &  $+$4.1&0.4   & 0.55     & 0.548&0.117 \\
9 Sgr A,B              & FH   &  60&4         &  $-$6.9&0.3   & 0.42     & 0.609&0.064 \\
$\alpha$ Cam           & C    &  63&5         &  $-$4.5&0.4   & 0.63     & 0.521&0.090 \\
$\zeta$ Oph            & FN   &  64&4         & $-$15.0&0.3   & 0.24     & 0.723&0.082 \\
CPD $-$59 2591         & F    &  67&5         &  $+$1.1&0.4   & 0.71     & 0.522&0.086 \\
HD 34\,656             & C    &  83&5         &  $+$5.4&0.4   & 0.64     & 0.625&0.096 \\
Herschel 36            & FN   &  84&4         &  $-$5.4&0.3   & 0.30     & 0.727&0.063 \\
V662 Car               & F    &  92&5         &  $-$6.3&0.4   & 0.85     & 0.478&0.054 \\
CE Cam                 & CH   &  98&4         &  $-$9.8&0.3   & 0.60     & 0.506&0.028 \\
HD 93\,162             & F    & 101&5         &  $-$9.2&0.4   & 0.98     & 0.553&0.063 \\
HD 192\,639            & NH   & 106&4         & $-$13.2&0.3   & 0.41     & 0.676&0.043 \\
HD 93\,250 A,B         & F    & 113&5         &  $-$5.8&0.4   & 0.78     & 0.537&0.054 \\
BD $-$16 4826          & N    & 115&5         &  $-$4.7&0.4   & 0.73     & 0.519&0.049 \\
HD 46\,150             & F    & 115&5         & $+$23.6&0.4   & 0.48     & 0.592&0.063 \\
CPD $-$59 2626 A,B     & F    & 117&5         &  $-$0.2&0.4   & 1.01     & 0.626&0.068 \\
HD 149\,452            & F    & 143&5         & $-$15.8&0.4   & 0.51     & 0.578&0.048 \\
HD 93\,129 Aa,Ab       & F    & 144&5         &  $-$7.5&0.4   & 1.08     & 0.594&0.050 \\
HD 93\,161 A           & F    & 145&5         &  $-$1.8&0.4   & 0.81     & 0.632&0.056 \\
HDE 326\,329           & F    & 149&5         &  $-$4.9&0.4   & 0.90     & 0.634&0.055 \\
HD 48\,279 A           & F    & 156&5         & $+$24.4&0.4   & 0.54     & 0.736&0.069 \\
HDE 322\,417           & F    & 157&5         & $-$12.9&0.4   & 0.87     & 0.602&0.048 \\
AE Aur                 & H    & 168&5         & $+$14.7&0.4   & 0.37     & 0.771&0.070 \\
CPD $-$47 2963 A,B     & F    & 169&5         & $+$19.8&0.4   & 0.29     & 0.771&0.070 \\
$\lambda$ Cep          & NH   & 170&4         & $-$15.1&0.3   & 0.87     & 0.649&0.025 \\
HD 199\,216            & H    & 172&5         & $-$12.3&0.4   & 0.29     & 0.795&0.073 \\
NGC 2024-1             & N    & 175&5         & $+$27.6&0.4   & 0.43     & 0.730&0.061 \\
ALS 15\,210            & F    & 177&5         & $-$12.5&0.4   & 0.79     & 0.658&0.050 \\
HD 46\,223             & H    & 177&5         & $+$25.1&0.4   & 0.37     & 0.711&0.057 \\
HDE 319\,702           & F    & 191&5         &  $-$5.4&0.4   & 0.48     & 0.665&0.047 \\
ALS 19\,613 A          & C    & 191&5         &  $-$3.9&0.4   & 0.57     & 0.621&0.041 \\
Cyg OB2-11             & N    & 199&5         & $-$10.4&0.4   & 0.58     & 0.588&0.036 \\
HDE 319\,703 A         & F    & 205&5         &  $-$4.8&0.4   & 0.72     & 0.654&0.042 \\
HT Sge                 & FH   & 206&4         &  $-$3.8&0.3   & 0.57     & 0.715&0.025 \\
ALS 2063               & F    & 209&5         & $+$12.3&0.4   & 1.13     & 0.564&0.032 \\
Cyg OB2-1 A            & N    & 209&5         &  $-$6.9&0.4   & 0.53     & 0.687&0.045 \\
Cyg OB2-15             & N    & 214&5         & $-$10.3&0.4   & 0.60     & 0.672&0.043 \\
ALS 4962               & F    & 218&5         &  $-$2.1&0.4   & 0.41     & 0.756&0.052 \\
Cyg OB2-20             & N    & 219&5         &  $-$7.3&0.4   & 0.61     & 0.685&0.043 \\
BD $+$36 4063          & N    & 222&5         & $-$14.0&0.4   & 0.64     & 0.706&0.045 \\
ALS 19\,693            & F    & 223&5         &  $-$5.9&0.4   & 0.49     & 0.746&0.050 \\
BD $-$13 4930          & F    & 225&5         &  $-$8.5&0.4   & 0.66     & 0.638&0.037 \\
HD 156\,738 A,B        & F    & 226&5         &  $-$5.7&0.4   & 0.50     & 0.661&0.039 \\
CPD $-$49 2322         & F    & 228&5         & $+$18.1&0.4   & 0.32     & 0.778&0.053 \\
BD $-$14 5014          & N    & 229&5         &  $-$5.2&0.4   & 1.04     & 0.639&0.036 \\
W 40 OS 1a             & C    & 233&5         &  $-$9.6&0.4   & 0.32     & 0.777&0.051 \\
NGC 1624-2             & N    & 234&5         &  $-$9.4&0.4   & 1.11     & 0.648&0.036 \\
\hline
\multicolumn{9}{l}{Codes: C, CARMENES; F, FEROS; N, FIES; H, HERMES.} \\
\end{tabular}
$\;\;$}
\label{KItable}
\addtolength{\tabcolsep}{3pt}
\end{table}

\addtocounter{table}{-1}

\begin{table}
\addtolength{\tabcolsep}{-3pt}
\caption{(continued).}
\centerline{$\;\;$
\begin{tabular}{lcr@{$\pm$}lr@{$\pm$}lcr@{$\pm$}l}
star                   & code & \mcii{EW}     & \mcii{$v$}    & W$_{90}$ & \mcii{$\frac{{\rm EW}_{7701.093}}{{\rm EW}_{7667.021}}$} \\
                       &      & \mcii{(m\AA)} & \mcii{(km/s)} & (\AA)    & \mcii{}                                                  \\
\hline
HD 207\,198            & CH   & 236&4         & $-$14.0&0.3   & 0.37     & 0.762&0.025 \\
BD $-$12 4979          & CF   & 236&4         &  $-$1.9&0.3   & 0.75     & 0.638&0.018 \\
Cyg OB2-B18            & H    & 236&5         & $-$12.3&0.4   & 0.45     & 0.771&0.050 \\
BD $+$60 513           & N    & 239&5         & $-$13.7&0.4   & 0.81     & 0.580&0.029 \\
HD 217\,086            & NH   & 240&4         & $-$19.5&0.3   & 0.67     & 0.617&0.016 \\
Cyg OB2-4 B            & N    & 242&5         &  $-$8.6&0.4   & 0.64     & 0.669&0.037 \\
BD $-$14 4922          & N    & 246&5         &  $+$2.4&0.4   & 0.92     & 0.589&0.029 \\
HDE 319\,703 Ba,Bb     & F    & 249&5         &  $-$3.6&0.4   & 0.67     & 0.637&0.033 \\
Tyc 8978-04440-1       & F    & 249&5         & $-$22.0&0.4   & 0.95     & 0.657&0.035 \\
HD 168\,076 A,B        & F    & 252&5         &  $-$5.7&0.4   & 0.72     & 0.663&0.035 \\
HD 168\,112 A,B        & FN   & 252&4         &  $-$0.1&0.3   & 0.83     & 0.664&0.018 \\
LS I $+$61 303         & N    & 252&5         & $-$13.4&0.4   & 1.28     & 0.688&0.038 \\
LS III $+$46 12        & NH   & 256&4         & $-$16.0&0.3   & 0.60     & 0.721&0.020 \\
HD 194\,649 A,B        & NH   & 259&4         & $-$17.4&0.3   & 0.98     & 0.648&0.016 \\
HDE 323\,110           & F    & 260&5         &  $-$8.4&0.4   & 0.56     & 0.660&0.034 \\
BD $-$14 5040          & N    & 263&5         &  $+$0.6&0.4   & 0.70     & 0.777&0.046 \\
MY Ser Aa,Ab           & F    & 264&5         &  $-$1.1&0.4   & 0.73     & 0.701&0.037 \\
Cyg OB2-8 A            & NH   & 267&4         &  $-$9.2&0.3   & 0.55     & 0.694&0.018 \\
BD $+$61 487           & N    & 268&5         & $-$19.4&0.4   & 1.43     & 0.619&0.029 \\
BD $-$11 4586          & CF   & 274&4         &  $-$4.5&0.3   & 0.75     & 0.655&0.016 \\
Cyg OB2-5 A,B          & NH   & 275&4         &  $-$9.6&0.3   & 0.45     & 0.758&0.021 \\
HD 168\,075            & FN   & 276&4         &  $-$5.2&0.3   & 0.74     & 0.686&0.017 \\
Tyc 7370-00460-1       & FN   & 279&4         &  $-$6.3&0.3   & 0.78     & 0.723&0.019 \\
Cyg OB2-7              & NH   & 280&4         &  $-$9.9&0.3   & 0.56     & 0.712&0.018 \\
V479 Sct               & FN   & 282&4         &  $+$2.4&0.3   & 0.61     & 0.722&0.018 \\
BD $-$13 4923          & FN   & 286&4         &  $-$3.1&0.3   & 0.85     & 0.629&0.014 \\
Sh 2-158 1             & NH   & 287&4         & $-$26.9&0.3   & 1.41     & 0.557&0.011 \\
HD 166\,734            & F    & 289&5         &  $-$6.8&0.4   & 0.67     & 0.742&0.038 \\
BD $+$62 2078          & N    & 292&5         & $-$16.0&0.4   & 0.62     & 0.729&0.036 \\
LS III $+$46 11        & NH   & 297&4         & $-$15.6&0.3   & 0.55     & 0.714&0.017 \\
ALS 18\,748            & F    & 301&5         & $-$22.7&0.4   & 1.08     & 0.623&0.026 \\
Pismis 24-1 A,B        & F    & 302&5         & $-$10.6&0.4   & 1.46     & 0.713&0.034 \\
Cyg OB2-A11            & NH   & 303&4         &  $-$9.3&0.3   & 0.46     & 0.796&0.021 \\
Cyg X-1                & N    & 308&5         & $-$12.0&0.4   & 0.75     & 0.698&0.032 \\
BD $-$13 4929          & F    & 310&5         &  $-$5.8&0.4   & 0.65     & 0.725&0.034 \\
ALS 15\,108 A,B        & H    & 315&5         & $-$10.6&0.4   & 0.42     & 0.814&0.042 \\
HD 15\,570             & CH   & 316&4         & $-$12.8&0.3   & 1.11     & 0.686&0.015 \\
ALS 19\,307            & N    & 316&5         &  $-$5.5&0.4   & 0.88     & 0.665&0.028 \\
Cyg OB2-3 A,B          & NH   & 317&4         &  $-$6.3&0.3   & 0.51     & 0.745&0.017 \\
ALS 15\,133            & N    & 319&5         &  $-$8.4&0.4   & 0.57     & 0.715&0.032 \\
Cyg OB2-73             & N    & 343&5         & $-$11.7&0.4   & 0.70     & 0.804&0.037 \\
NGC 3603 HST-5         & F    & 343&5         &  $-$9.6&0.4   & 0.70     & 0.703&0.029 \\
Cyg OB2-22 A           & H    & 344&5         &  $-$9.9&0.4   & 0.45     & 0.765&0.034 \\
Cyg OB2-B17            & N    & 348&5         &  $-$8.4&0.4   & 0.57     & 0.762&0.033 \\
THA 35-II-42           & F    & 350&5         &  $+$3.0&0.4   & 0.54     & 0.774&0.034 \\
ALS 15\,131            & N    & 354&5         &  $-$7.8&0.4   & 0.66     & 0.845&0.040 \\
HD 17\,603             & CH   & 355&4         & $-$22.5&0.3   & 1.23     & 0.666&0.013 \\
Bajamar star           & CNH  & 357&3         & $-$12.5&0.3   & 0.46     & 0.818&0.013 \\
Cyg OB2-22 C           & N    & 362&5         &  $-$9.1&0.4   & 0.65     & 0.716&0.028 \\
BD $+$66 1674          & NH   & 370&4         & $-$16.5&0.3   & 0.51     & 0.769&0.016 \\
Cyg OB2-27 A,B         & N    & 378&5         &  $-$9.3&0.4   & 0.73     & 0.704&0.026 \\
BD $+$66 1675          & NH   & 385&4         & $-$16.6&0.3   & 0.52     & 0.777&0.016 \\
Cyg OB2-12             & CH   & 387&4         &  $-$9.0&0.3   & 0.50     & 0.809&0.017 \\
BD $-$13 4927          & F    & 388&5         &  $-$3.6&0.4   & 0.61     & 0.763&0.030 \\
Cyg OB2-9              & N    & 409&5         &  $-$9.7&0.4   & 0.58     & 0.729&0.026 \\
V889 Cen               & F    & 413&5         & $-$29.3&0.4   & 1.32     & 0.680&0.023 \\
Tyc 4026-00424-1       & CH   & 416&4         & $-$16.9&0.3   & 0.50     & 0.843&0.017 \\
HDE 326\,775           & F    & 420&5         & $-$21.2&0.4   & 0.97     & 0.670&0.022 \\
V747 Cep               & NH   & 424&4         & $-$17.3&0.3   & 0.58     & 0.758&0.013 \\
KM Cas                 & N    & 429&5         & $-$24.0&0.4   & 1.52     & 0.699&0.023 \\
\hline
\multicolumn{9}{l}{Codes: C, CARMENES; F, FEROS; N, FIES; H, HERMES.} \\
\end{tabular}
}
\addtolength{\tabcolsep}{3pt}
\end{table}

$\,\!$\indent We list in Table~\ref{ISMlines} the relevant ISM lines that we have found in the 7400--8000~\AA\ range in our spectra. The most significant contaminant for the measurement
of the broad absorption band at 7700~\AA\ is the \KId{7667.021,7701.093} doublet. We have measured those two lines in the combined LiLiMaRlin spectra for each star and telescope 
(as mentioned in the text, different epochs of the same star with the same telescope are used when available to alleviate the effect of telluric lines) and the results are given
in Table~\ref{KItable}. Given the variety of data and the need to accurately eliminate telluric lines, each line has been measured with an interactive program that allows us to
set the continuum and integrating regions. No profiles have been fit, as many stars show multiple kinematic components. Instead, the lines have been integrated to derive their
equivalent widths (EWs, used for sorting in the table), central velocities ($v$), and width that contains 90\% of the flux (W$_{90}$). The latter value is selected because, given 
the complexity of the profiles, it captures the width of the line better than smaller percentages and higher ones may be subject to larger systematic effects for low values of 
the EW.  In Table~\ref{KItable} we show the values for \KI{7701.093} and the equivalent width ratio with \KI{7667.021} instead of the opposite for two reasons. First, 
\KI{7701.093} is the weaker line (in the high S/N, low intensity regime the equivalent width ratio should be 0.33) so it saturates later and choosing it as the reference allows for a 
larger dynamic range in the EW.  Second and most importantly, \KI{7667.021} is more affected by O$_2$ telluric lines and is, therefore, more prone to systematic errors. 
The uncertainties in all quantities have been calculated from the comparison of the measurements of the same star using different telescopes and they encompass both random and 
systematic effects but the latter are the dominant ones. When in doubt, we have erred on the side of caution e.g. the uncertainties in the EW are likely overestimated for the stars 
with low values. For the sample in common with \citet{WeltHobb01} we find a good agreement for the EWs.
The value of W$_{90}$ for measurements with FEROS (code F in the table) and FIES (code N) have been reduced by 0.05~\AA\ and 0.10~\AA, respectively, 
to account for the lower spectral resolution of those spectrographs. The correction values were determined empirically by comparing stars observed with those spectrographs and
with the higher resolution CARMENES (code C) or HERMES (code H).
%, or UVES (code U). 

%https://www.nist.gov/system/files/documents/srd/jpcrd37200896p.pdf 0.33 = (3.75e7/3.80e7)*1/3

\begin{figure}
\centerline{\includegraphics[width=\linewidth]{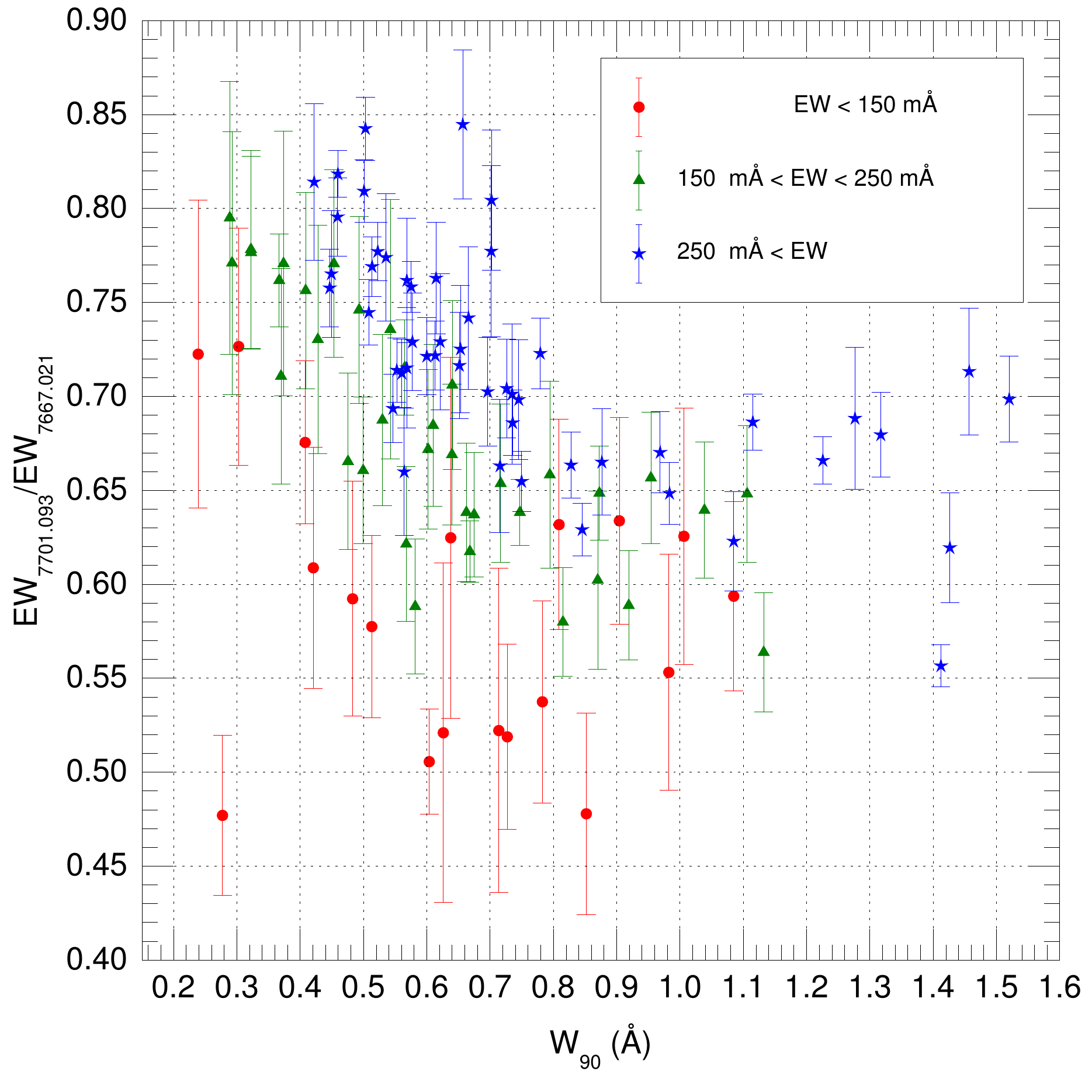}}
\caption{Equivalent-width ratio for the K\,{\sc i} doublet as a function of W$_{90}$ for \KI{7701.093}. Symbols are colour-coded by equivalent width. Objects with uncertainties in 
         the vertical axis greater than 0.1 are not plotted.}
\label{KI_W90_EWrat}
\end{figure}

\begin{figure}
\centerline{\includegraphics[width=\linewidth]{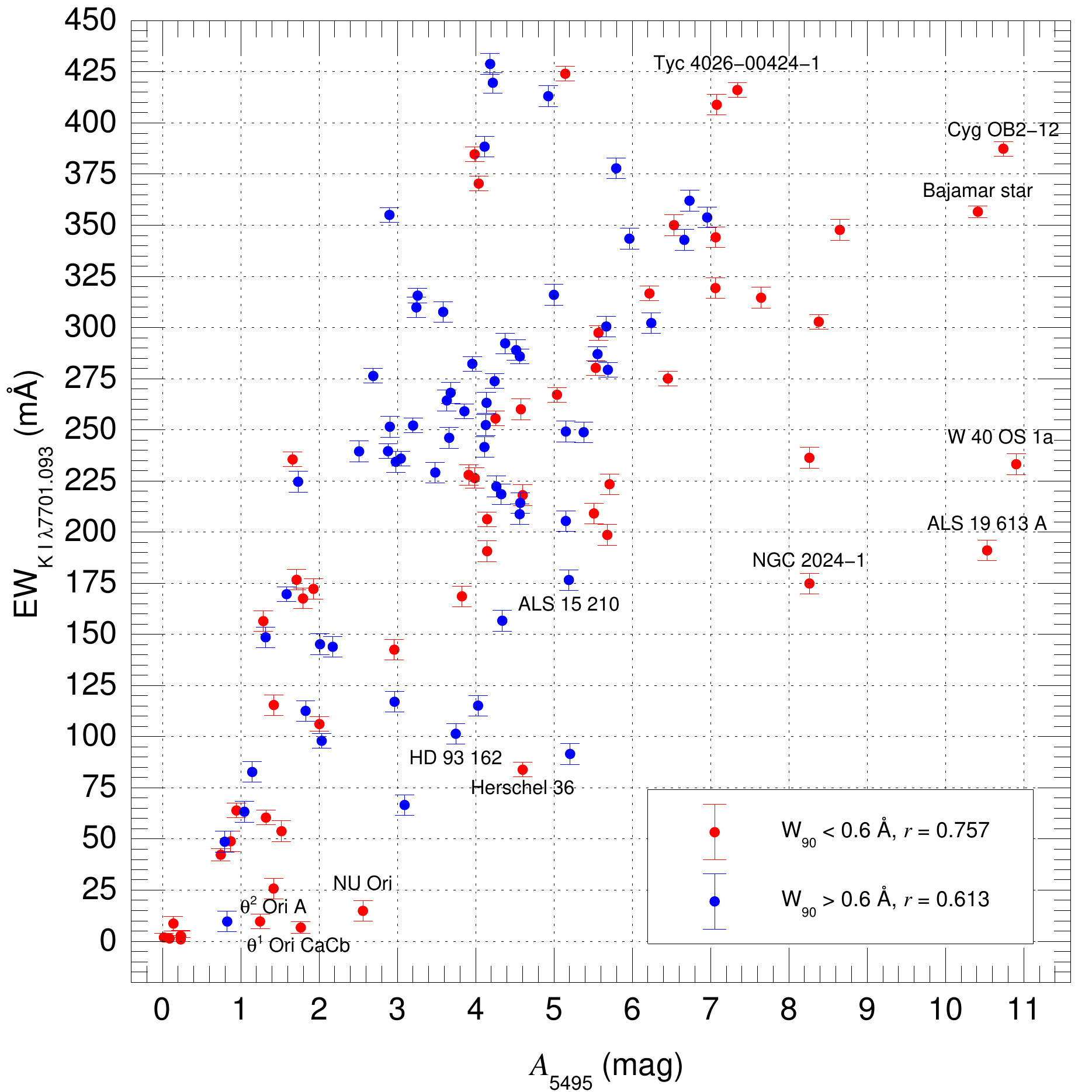}}
 \caption{\EWKIr\ as a function of \AV. Symbols are colour-coded by W$_{90}$ for \KI{7701.093}. Pearson correlation coefficients for the two subsets are given in
          the legend. Some targets discussed in the text are labelled.}
\label{av_KIr}
\end{figure}

Table~\ref{KItable} shows a large range of \EWKIr, from difficult to detect to more than 0.4~\AA. In Fig.~\ref{KI_W90_EWrat} we show the relationship between \EWKIr, W$_{90}$
for \KI{7701.093}, and the EW ratio. As \EWKIr\ increases, the points move towards the upper right as a result of the increase in saturation of the lines. On the one hand, the 
flattening at the bottom of the line increases its width and, on the other hand, saturation appears first in \KI{7667.021} and increases the EW ratio (a curve of growth effect). 
Within a colour group in Fig.~\ref{KItable}, the range of values in W$_{90}$ corresponds to the different kinematics of the intervening gas: sightlines with single components are 
located towards the left and those with multiple components towards the right. This has two consequences: the same \EWKIr\ does not correspond to the same amount of intervening 
material (which is larger for smaller W$_{90}$ at constant \EWKIr) and the EW ratio increases for smaller W$_{90}$, as the two lines are moved towards a flatter part of the curve 
of growth in the absence of multiple components. 

In Fig.~\ref{av_KIr} we compare \EWKIr\ with the extinction \AV\ (see main text for its calculation). The two quantities are correlated but there is a significant spread. 
If we restrict the sample to those objects with ``narrow'' profiles (W$_{90} <$~0.6~\AA) the correlation improves, as for wider profiles \EWKIr\ can depend heavily
on the particular gas kinematics (many of those sightlines have multiple kinematic components, significantly those in the Carina nebula, see \citealt{Walb82d}). However, variations in 
the column density ratios of dust and K\,{\sc i} must also exist, as evident from the existence of some sightlines prominently poor in K\,{\sc i}. Several of those cases are in \HII\ regions
so we are likely observing an ionization effect \citep{WeltHobb01}, as the ionization potential of potassium is low.
In general, \EWKIr\ is not a very good predictor of extinction, especially if the line is narrow. In our sample we have stars 
with \EWKIr\ of $\approx$~200~m\AA\ with \AV\ from less than 2~mag to more than 10~mag. Conversely, \AV\ is not a good predictor of the \EWKIr, especially if the line is wide.

\begin{table}
\caption{EWs for the $Q$(2) (3,0) C$_2$ Phillips band line at 7724.221~\AA.}
\centerline{
\begin{tabular}{lcr@{$\pm$}ll}
star                  & code & \mcii{EW}     & notes                  \\
                      &      & \mcii{(m\AA)} &                        \\
\hline
AE Aur                & C    &  1.8&1.2      & (2,0) detection by F94 \\
ALS 15\,210           & F    &  2.2&1.2      &                        \\
BD $+$66 1674         & NH   &  8.6&0.8      &                        \\
BD $+$66 1675         & NH   &  8.9&0.8      & (2,0) detection by v89 \\
Bajamar star          & CNH  &  9.4&0.7      &                        \\
CPD $-$47 2963 A,B    & F    &  2.2&1.2      &                        \\
Cyg OB2-12            & CH   &  9.0&0.8      & detections by v89, G94 \\
Cyg OB2-5 A,B         & H    &  4.4&1.2      & (1,0) detection by G94 \\
Cyg OB2-B17           & N    &  9.1&1.2      &                        \\
HD 156\,738 A,B       & F    &  2.3&1.2      & not detected by v89    \\
HD 207\,198           & CH   &  0.9&0.8      & (2,0) detection by v89 \\
LS III $+$46 11       & H    &  6.4&1.2      &                        \\
NGC 2024-1            & C    &  8.5&1.2      &                        \\
Tyc 4026-00424-1      & CH   &  9.3&0.8      &                        \\
V747 Cep              & NH   &  9.0&0.8      &                        \\
W 40 OS 1a            & C    & 19.6&1.2      &                        \\
$\lambda$ Cep         & N    &  0.6&0.6      & (2,0) detection by v89 \\
$\zeta$ Oph           & NH   &  0.5&0.5      & (2,0) detection by v89 \\
\hline
\multicolumn{5}{l}{Codes: C, CARMENES; F, FEROS; N, FIES; H, HERMES.} \\
\multicolumn{5}{l}{v89: \citet{vanDBlac89}. F94: \citet{Fedeetal94}.} \\
\multicolumn{5}{l}{G94: \citet{GredMunc94}.                         } \\
\end{tabular}
$\;\;$}
\label{C2table}
\end{table}

\begin{figure}
\centerline{\includegraphics[width=\linewidth]{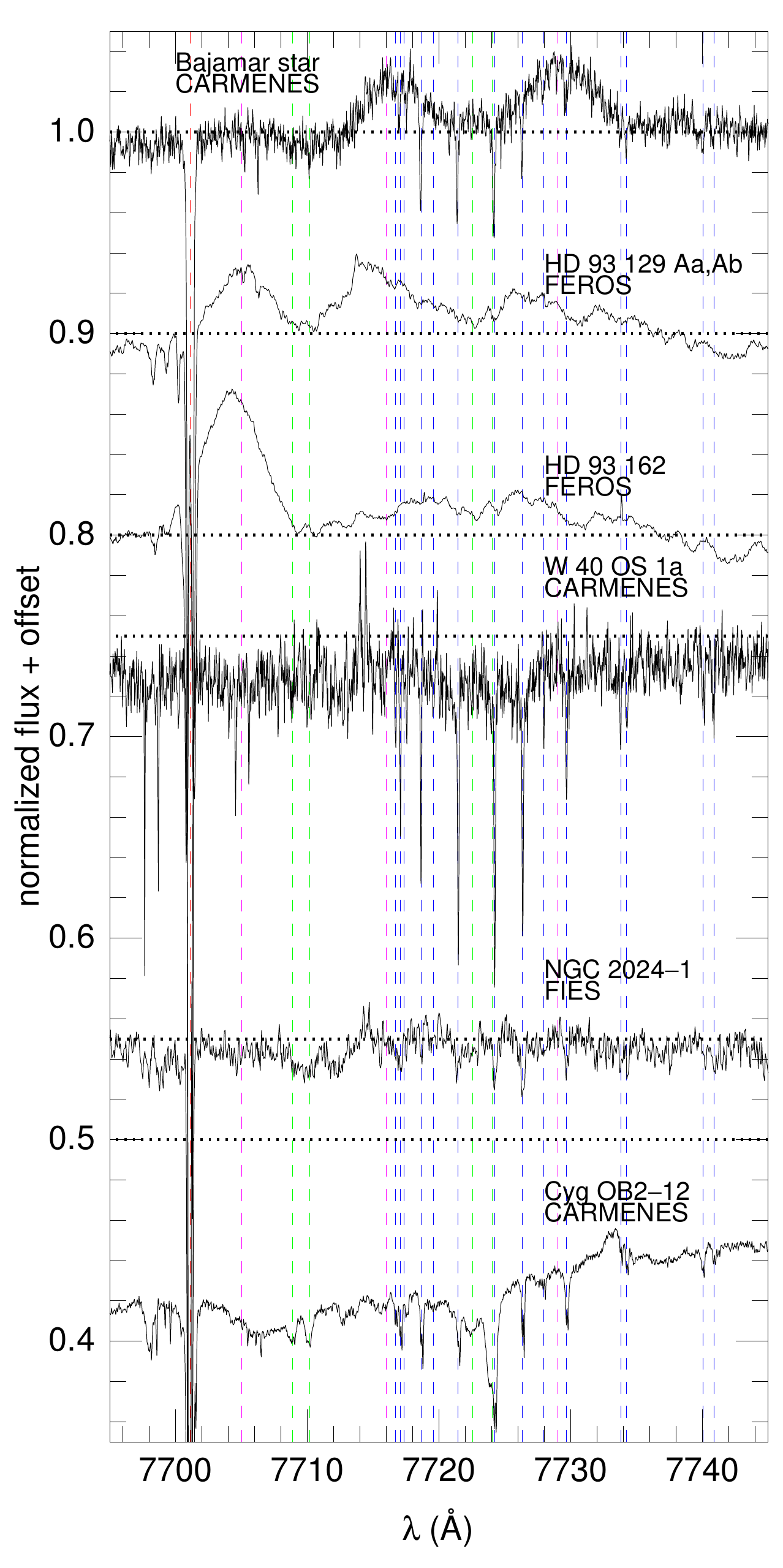}}
 \caption{Sample \'echelle spectra to showcase the rich ISM and stellar spectra in the 7695--7745~\AA\ region. Red, blue, green, and magenta mark the position of atomic, molecular,
          DIB, and stellar lines, respectively, the first three (ISM) lines in absorption and the last (stellar) in emission. The thick dashed horizontal lines mark the position of
          the continua (note that Cyg OB2-12 has a deep 7700~\AA\ absorption band, see Fig.~\ref{spectra_HR}). All spectra have been shifted to the \KI{7701.093} (which is deep in 
          all cases) but note that the Carina stars have complex profiles.}
\label{spectra_lines}
\end{figure}

Another ISM contaminant in our spectra is the (3,0) Phillips band of C$_2$ \citep{vanDBlac84,vanDBlac86,Kazmetal10b,IglG11}. 
The band is much weaker than the \KId{7667.021,7701.093} doublet and, indeed, 
it can only be clearly seen in a few objects after subtracting the (weak) telluric O$_2$ lines in that wavelength region. For some stars the C$_2$ lines are seen superimposed on stellar
emission lines and weak DIBs (see below), which are much broader than the components of the molecular band. Except maybe for W40~OS~1a, the (3,0) Phillips band 
is not strong enough to make a significant addition to the measured EW of the broad absorption band studied in this paper. For the purpose of using it as a reference for the presence 
of C$_2$ in the line of sight we have only measured the EW for the $Q(2)$~$\lambda$7724.221 line (Table~\ref{C2table}), whose value is about one quarter of the total for the band 
(see e.g. Fig.~1 in \citealt{vanDBlac86}). We suspect the C$_2$ band is present in some of the other stars in 
the sample other than those listed in Table~\ref{C2table} but close to the noise level. We have chosen not to list those one-to-two sigma detections there except for the objects that had been
previously detected.

In addition to the K\,{\sc i} doublet and the C$_2$ band, we have also searched for DIBs in the region that could be strong enough to interfere with our measurement of the broad absorption 
band. There are several listed in the literature but the majority of them are too weak to produce a significant relative contribution. Two of the DIBs listed by \citet{JennDese94}
are strong enough to have significantly large EWs and they are centred at 7434.1$\pm$0.2~\AA\ and 7929.9$\pm$0.1~\AA, respectively. 
In those two cases we have fitted Gaussian profiles to our high resolution spectra for the stars where they are detected, which are the majority of our sample. Assuming they are produced at 
the same velocity as the K\,{\sc i} doublet to correct for their Doppler shifts, we have combined the results to obtain the average properties of the two DIBs  
(Table~\ref{ISMlines}). In both cases we obtain FWHMs slightly smaller than \citet{JennDese94} but similar central wavelengths (converting our results to air wavelengths). In any
circumstance, both DIBs are located at the edges of the region analysed, far from where they can interfere with our measurements of the broad absorption band, so they can be ignored for
its analysis. As discussed in the text, a different issue is that of the DIB listed by \citet{JennDese94} as having a central wavelength of 7709.7~\AA\ (air) and a FWHM of 33.5~\AA, as we 
think that DIB is just the central region of the broad absorption band.

There are two additional DIBs around 7709--7710~\AA\ and another two around 7722--7724~\AA\ that are narrower and that deserve our attention for a different reason. They are located 
within the broad absorption band but their EWs are too small to make a significant dent in its measurement. However, they interfere with three stellar emission lines (and with some C$_2$ 
absorption lines), as they are located in 
the valleys between them. We have measured their properties to ensure that their contribution to them is small and we list their central wavelengths and FWHM in Table~\ref{ISMlines}. The first
two DIBs are in the \citet{Hobbetal08} list and the last two in the \citet{JennDese94} compilation. The central wavelengths and FWHM we measure for the two DIBs around 7709--7710~\AA\ are 
essentially the same (once they are converted from vacuum wavelengths to air ones) as the ones measured by \citet{Hobbetal08} for a single star (HD~204\,827). The same is true for the comparison
between the 7724.03~\AA\ DIB and its \citet{JennDese94} equivalent but for the last DIB (the one centred at 7722.58~\AA) we obtain a significantly larger FWHM.

\section{Stellar lines in the 7400--8000~\AA\ range}

\begin{table}
\caption{OB stellar lines in the 7400--8000~\AA\ region (vacuum wavelengths) discussed in this paper.}
\label{stellarlines}
\centerline{
\begin{tabular}{lr@{.}lcll}
ion          & \mcii{$\lambda_{\rm c}$} & type$\dagger$  & stellar type       & notes               \\
             & \mcii{($\AA$)}           &                &                    &                     \\
\hline
He\,{\sc ii} & 7594&844                 & abs            & O                  & strong              \\
N\,{\sc iv}? & \mlii{7705}              & em             & early O SGs        & FWHM $\approx$5 \AA \\
O\,{\sc iv}? & \mlii{7716}              & em             & early/mid O        & FWHM $\approx$6 \AA \\
C\,{\sc iv}? & \mlii{7729}              & em             & early/mid O        & FWHM $\approx$5 \AA \\
O\,{\sc i}   & 7774&083                 & abs            & late B/early A SGs & strong triplet      \\
O\,{\sc i}   & 7776&305                 & abs            & late B/early A SGs & strong triplet      \\
O\,{\sc i}   & 7777&527                 & abs            & late B/early A SGs & strong triplet      \\
\hline
\multicolumn{6}{l}{$\dagger$ abs: absorption, em: emission.} \\
\end{tabular}
}
\end{table}

\begin{table*}
\caption{Results for the stellar emission lines. Stars are sorted by supergiants/non-supergiants first and by spectral subtype second.}
\centerline{
\begin{tabular}{llcr@{$\pm$}lcr@{$\pm$}lcr@{$\pm$}l}
star                  & spectral type          & $\lambda$7705 & \mcii{EW$_{\lambda 7705}$} & $\lambda$7716 & \mcii{EW$_{\lambda 7716}$} & $\lambda$7729 & \mcii{EW$_{\lambda 7729}$} \\
                      &                        & code          & \mcii{(\AA)}               & code          & \mcii{(\AA)}               & code          & \mcii{(\AA)}               \\
\hline
HD 93\,129 Aa,Ab      & O2 If* + O2 If* +OB?   & F             & 0.18&0.01                  & F             & 0.12&0.02                  & F             & 0.09&0.02                  \\
THA 35-II-42          & O2 If*/WN5             & F             & 0.21&0.01                  & F             & 0.10&0.02                  & F             & 0.09&0.02                  \\
HD 93\,162            & O2.5 If*/WN6 + OB      & F             & 0.38&0.01                  & F             & 0.10&0.02                  & F             & 0.10&0.02                  \\
Cyg OB2-22 A          & O3 If*                 & H             & 0.13&0.01                  & H             & 0.17&0.02                  & H             & 0.13&0.02                  \\
Cyg OB2-7             & O3 If*                 & NH            & 0.22&0.01                  & NH            & 0.05&0.02                  & NH            & 0.08&0.01                  \\
LS III $+$46 11       & O3.5 If* + O3.5 If*    & ---           & \mcii{---}                 & H             & 0.11&0.02                  & H             & 0.14&0.02                  \\
ALS 15\,210           & O3.5 If* Nwk           & F             & 0.09&0.01                  & F             & 0.07&0.02                  & F             & 0.12&0.02                  \\
Pismis 24-1 A,B       & O3.5 If*               & FH            & 0.07&0.01                  & FH            & 0.16&0.02                  & FH            & 0.15&0.01                  \\
HD 15\,570            & O4 If                  & CH            & 0.09&0.01                  & CH            & 0.10&0.02                  & CH            & 0.13&0.01                  \\
CPD $-$47 2963 A,B    & O5 Ifc                 & ---           & \mcii{---}                 & F             & 0.06&0.02                  & F             & 0.04&0.02                  \\
ALS 4962              & ON5 Ifp                & F             & 0.08&0.01                  & F             & 0.12&0.02                  & F             & 0.06&0.02                  \\
ALS 2063              & O5 Ifp                 & F             & 0.07&0.01                  & F             & 0.04&0.02                  & F             & 0.11&0.02                  \\
HD 192\,639           & O7.5 Iabf              & ---           & \mcii{---}                 & NH            & 0.08&0.02                  & NH            & 0.06&0.01                  \\
HD 34\,656            & O7.5 II(f)             & ---           & \mcii{---}                 & C             & 0.02&0.02                  & C             & 0.02&0.02                  \\
ALS 19\,613 A         & O2/4 Vp                & ---           & \mcii{---}                 & CN            & 0.29&0.02                  & CN            & 0.24&0.01                  \\
Sh 2-158 1            & O3.5 V((fc)) + O9.5: V & ---           & \mcii{---}                 & H             & 0.17&0.02                  & H             & 0.10&0.02                  \\
Bajamar star          & O3.5 III(f*) + O8:     & CH            & 0.02&0.01                  & CNH           & 0.16&0.01                  & CNH           & 0.18&0.01                  \\
HD 93\,205            & O3.5 V((f)) + O8 V     & F             & 0.01&0.01                  & F             & 0.21&0.02                  & F             & 0.17&0.02                  \\
HD 46\,223            & O4 V((f))              & C             & 0.04&0.01                  & C             & 0.06&0.02                  & C             & 0.10&0.02                  \\
HD 168\,076 A,B       & O4 IV(f)               & F             & 0.06&0.01                  & F             & 0.10&0.02                  & F             & 0.10&0.02                  \\
9 Sgr A,B             & O4 V((f))              & FH            & 0.05&0.01                  & FH            & 0.10&0.02                  & FH            & 0.14&0.01                  \\
HD 93\,250 A,B        & O4 IV(fc)              & F             & 0.03&0.01                  & F             & 0.12&0.02                  & F             & 0.13&0.02                  \\
BD $-$13 4923         & O4 V((f)) + O7.5 V     & ---           & \mcii{---}                 & FN            & 0.07&0.02                  & FN            & 0.07&0.01                  \\
LS III $+$46 12       & O4.5 IV(f)             & H             & 0.09&0.01                  & NH            & 0.16&0.02                  & NH            & 0.14&0.01                  \\
HD 168\,112 A,B       & O5 IV(f) + O6: IV:     & ---           & \mcii{---}                 & F             & 0.07&0.02                  & F             & 0.04&0.02                  \\
HD 46\,150            & O5 V((f))z             & C             & 0.03&0.01                  & CF            & 0.06&0.02                  & CF            & 0.10&0.01                  \\
BD $-$16 4826         & O5.5 V((f))z           & ---           & \mcii{---}                 & N             & 0.10&0.02                  & N             & 0.14&0.02                  \\
$\theta^{1}$ Ori Ca,C & O7 f?p var             & ---           & \mcii{---}                 & CH            & 0.04&0.02                  & CH            & 0.05&0.01                  \\
15 Mon Aa,Ab          & O7 V((f))z + B1: Vn    & ---           & \mcii{---}                 & C             & 0.04&0.02                  & C             & 0.02&0.02                  \\
$\lambda$ Ori A       & O8 III((f))            & ---           & \mcii{---}                 & C             & 0.06&0.02                  & C             & 0.02&0.02                  \\
\hline
\multicolumn{8}{l}{Codes: C, CARMENES; F, FEROS; N, FIES; H, HERMES.} \\
\end{tabular}
$\;\;$}
\label{emissiontable}
\end{table*}

$\,\!$\indent We list in Table~\ref{stellarlines} the most relevant stellar lines that we have found in the 7400--8000~\AA\ range in the OB stars in our sample. As discussed in the text,
the main goal in this paper is to accurately measure the broad interstellar absorption band and to fulfil it we need to include the effect of the stellar lines in our spectra as
measured in the high-resolution spectra, as the lines are difficult or impossible to identify in the low-resolution STIS data. Nevertheless, in this Appendix we also list some of
the characteristics of the stellar lines for their intrinsic interest, as this is a relatively unexplored part of the optical spectrum for OBA stars. 

The two strongest multiplets detected in our spectra are the \HeII{7594.844} singlet and the \OIt{7774.083,7776.305,7777.527} triplet, both seen in absorption, the first one for O stars
and the second one for late B/early A supergiants. The \HeII{7594.844} line is difficult to see in ground-based data because it is located close to the head of the O$_2$ telluric
band. The \OIt{7774.083,7776.305,7777.527} triplet, on the other hand, is easy to detect in our ground-based spectra. In both cases they are located away from the centre of the
absorption band and they do not interfere much with its measurement. Also, both lines have well-studied equivalents in other parts of the spectrum and, therefore, do not provide
much new information about their stars.

We detect three broad emission lines in the spectra of early- and mid-O stars in the 7700-7735 region with approximate central wavelengths of 7705~\AA, 7716~\AA, and 7729~\AA, 
respectively (Fig.~\ref{spectra_lines}). The determination of the precise central wavelengths is complicated because broad emission lines in O stars originate 
in strong winds and their value may depend on the specific circumstances of each star. For a significant fraction of the stars with emission lines we only have FEROS spectroscopy, which
typically has residual effects from the flat-field correction left from the pipeline processing. We have measured the equivalent widths of the three lines for the objects that show them
and list the results is Table~\ref{emissiontable}.

The behaviour of the 7705~\AA\ line differs significantly from those of the other two. It appears only in spectral types O5 and earlier and is strong in supergiants and weak in objects of
lower luminosity classes. The object with the strongest emission in our sample is HD~93\,162, a ``hot-slash'' star \citep{Sotaetal14}, see Fig.~\ref{spectra_lines}. We have searched
in the Atomic Line List\footnote{\url{http://www.pa.uky.edu/~peter/atomic/}} and the most likely identification for the line is that it originates in N\,{\sc iv}.

The other two lines have similar behaviours, appearing in spectral subtypes from O2 to O8 (stronger in earlier subtypes) and with no strong dependence on luminosity class. However, given 
the relatively low S/N of some of the spectra and the issues with FEROS data previously mentioned, that statement should be taken with care. In some cases (e.g. Bajamar star, 
Fig.~\ref{spectra_lines}) the two lines are strong, similar in intensity, and with well-defined central wavelengths but in others it may be possible that the lines are actually more than
two. We have searched in the Atomic Line List and our best candidate for the 7716~\AA\ line is a couple of O\,{\sc iv} lines while for the 7729~\AA\ line we have found a group of 
C\,{\sc iv} lines in the 7726-7730~\AA\ range. Other authors have identified the latter line as originating in C\,{\sc iv} in WR stars \citep{Calletal20}. Better data are needed to
provide a more thorough identification.

\bsp	% typesetting comment
\label{lastpage}
\end{document}